\documentclass[format=acmsmall, screen=true, dvipsnames]{acmart}

\newcommand{\oj}[1]{\textbf{#1}}

\usepackage{graphicx}
\usepackage[space]{grffile}
\usepackage{latexsym}
\usepackage{textcomp}
\usepackage{longtable}
\usepackage{multirow,booktabs}
\usepackage{amsfonts,amsmath,amssymb}
\usepackage{natbib}
\usepackage{url}
\usepackage[normalem]{ulem}
\hypersetup{colorlinks=false,pdfborder={0 0 0}}
\newif\iflatexml\latexmlfalse
\DeclareGraphicsExtensions{.pdf,.PDF,.png,.PNG,.jpg,.JPG,.jpeg,.JPEG}

\usepackage[utf8]{inputenc}
\usepackage[T2A]{fontenc}
\usepackage[ngerman,polish,english]{babel}

\acmJournal{CSUR}

\begin{document}

\citestyle{acmauthoryear}
\setcopyright{acmcopyright}

\title[A Survey on Online Judge Systems]{A Survey on Online Judge Systems and Their Applications}

\author{Szymon Wasik}
\authornote{This is the corresponding author}
\authornote{SW and MA contributed equally to the paper}
\orcid{0000-0002-2753-2862}
\affiliation{%
   \institution{Poznan University of Technology}
   \department{Institute of Computing Science}
   \streetaddress{Piotrowo 2}
   \city{Poznan}
   \postcode{60-965}
   \country{Poland}}
\affiliation{%
   \institution{Polish Academy of Sciences}
   \department{Institute of Bioorganic Chemistry}
   \streetaddress{Z. Noskowskiego 12/14}
   \city{Poznan}
   \postcode{61-704}
   \country{Poland}}
\email{szymon.wasik@cs.put.poznan.pl}

\author{Maciej Antczak}
\orcid{0000-0002-5320-2023}
\affiliation{%
   \institution{Poznan University of Technology}
   \department{Institute of Computing Science}
   \streetaddress{Piotrowo 2}
   \city{Poznan}
   \postcode{60-965}
   \country{Poland}}
\email{maciej.antczak@cs.put.poznan.pl}   
   
\author{Jan Badura}
\affiliation{%
   \institution{Poznan University of Technology}
   \department{Institute of Computing Science}
   \streetaddress{Piotrowo 2}
   \city{Poznan}
   \postcode{60-965}
   \country{Poland}}
\email{jan.badura@cs.put.poznan.pl}
   
\author{Artur Laskowski}
\affiliation{%
   \institution{Poznan University of Technology}
   \department{Institute of Computing Science}
   \streetaddress{Piotrowo 2}
   \city{Poznan}
   \postcode{60-965}
   \country{Poland}}
\email{artur.laskowski@cs.put.poznan.pl}   
   
\author{Tomasz Sternal}
\affiliation{%
   \institution{Poznan University of Technology}
   \department{Institute of Computing Science}
   \streetaddress{Piotrowo 2}
   \city{Poznan}
   \postcode{60-965}
   \country{Poland}}
\email{tomasz.m.sternal@student.put.poznan.pl}

\begin{abstract}
Online judges are systems designed for the reliable evaluation of algorithm source code submitted by users, which is next compiled and tested in a homogeneous environment. Online judges are becoming popular in various applications. Thus, we would like to review the state of the art for these systems. We classify them according to their principal objectives into systems supporting organization of competitive programming contests, enhancing education and recruitment processes, facilitating the solving of data mining challenges, online compilers and development platforms integrated as components of other custom systems. Moreover, we introduce a formal definition of an online judge system and summarize the common evaluation methodology supported by such systems. Finally, we briefly discuss an Optil.io platform as an example of an online judge system, which has been proposed for the solving of complex optimization problems. We also analyze the competition results conducted using this platform. The competition proved that online judge systems, strengthened by crowdsourcing concepts, can be successfully applied to accurately and efficiently solve complex industrial- and science-driven challenges. 
\end{abstract}%



\thanks{All authors were supported by the National Center for Research and Development, Poland [grant no. LIDER/004/103/L-5/13/NCBR/2014]. Moreover, development tools JIRA and Bitbucket were shared by PLGrid infrastrucutre. The authors would like to thank Szymon Acedanski from Warsaw University for his advice on the implementation of execution time measurements methods.}

\maketitle

\renewcommand{\shortauthors}{S. Wasik et al.}

\section{Introduction}

In 1970, when Texas A\&M University organized the first edition of the ACM International Collegiate Programming Contest (ICPC), no one could have guessed that, in a few dozen years, it would be the largest and most prestigious programming contest in the world. In 2015, over 40,000 students from almost 3,000 universities and 102 countries participated in a regional phase of this contest \citep{ICPC_2016}. During the contest, lasting 5 hours, participants solve from 8 to 13 algorithmic problems. The winner is the team that first solves the highest number of problems. The key component of this contest environment is a system that automatically verifies the correctness of solutions submitted by participants. It assesses the correctness of the submitted solutions based on the results obtained from their execution on predefined test sets. It also verifies that the solution does not exceed resource utilization limits (such as time and memory). Based on the conducted evaluation, the online ranking of all participants is computed and presented in real-time during the ongoing contest.

Since the first finals of the ICPC, many other algorithmic competitions requiring similar automatic evaluation systems have been started. Most likely, the most important of them is the International Olympiad in Informatics (IOI), first organized in 1989. It is comparable to the ICPC but dedicated to secondary school pupils. The biggest differences between the aforementioned competitions are the following: the participants solve problems individually instead of collaboratively in teams; the number of problems considered by the IOI is lower than that by the ICPC; and in the IOI, the participants receive partial scores for each submitted solution instead of the binary scoring applied in the ICPC (every solution is treated as only correct or incorrect). Detailed information regarding the basic rules and scoring functions used by the most popular programming competitions was published by Cormack et al. \citep{cormack2006structure}.

The important supplemental role of the IOI is as a place for building an international community of people committed to the organization of programming contests. One of its important activities was foundation of the \textit{Olympiads in Informatics} journal, which publishes papers submitted annually for presentation in the conference organized in parallel to the IOI \citep{Dagiene_2007}. It provides an international forum for discussion regarding the experiences gained during national and international Olympiads in Informatics, including preparation of interesting problems and software supporting their organization. As mentioned before, the most important entries among such software are systems that automatically evaluate solutions submitted by participants; such systems are called \textit{online judges}. The term \textit{online judge} was introduced for the first time by Kurnia, Lim, and Cheang in 2001 as an online platform that supports fully-automated, real-time evaluation of source code, binaries, or even textual output submitted by participants competing in a particular challenge \citep{Kurnia_2001}. However, the development of online judge systems boasts a much longer history, dating back to 1961 when they emerged at Stanford University \citep{Leal98automaticgrading,Forsythe_1965}.

During the design and implementation of online judge systems, many important factors should be taken into consideration. The key issue is the security of such a system. The concept of an online judge assumes that the user submits the solution as the source code, or sometimes even the executable file, which will be evaluated in the next step, often in a cloud-based infrastructure. The designers of the online judge should ensure that the system is resistant to a broad range of attacks, such as forcing a high compilation time, modifying the testing environment or accessing restricted resources during the solution evaluation process. A detailed description of possible attack types and the methods of protection used against them in 2006 during Slovakian programming competitions can be found in \citep{Forivsek_2006a}. Unfortunately, the solutions of security threats presented in the aforementioned paper are mostly outdated, although their causes are still unchanged. Currently, the most popular methods of avoidance of such issues rely on the execution of submitted solutions in dedicated sandboxes managed by the online judge system  \citep{Yi_2014}, such as virtualization, LXC containers \citep{Felter_2015}, and the Docker framework \citep{merkel2014docker}. Such approaches could significantly increase the safety and reliability of the system.

Another important aspect that should be taken into consideration during the development of such systems is the measurement precision of execution time. The time limit for a single test case execution is often measured in milliseconds, and thus the performance analysis method used during evaluation should be sufficiently sensitive and deterministic to precisely distinguish such small fractions of time and ensure reproducible measurements of consecutive executions of the same code for the particular test case. There are various methods that can be used to measure the processing time, such as the utilization of simple command line utilities, analysis of hardware performance counters, code instrumentation or even code sampling. Applications of each entail various advantages, as well as disadvantages, that one should be aware of related to measurement precision, time overhead and available integration methods \citep{Ilsche_2015}.

In all online judge systems, one of the requirements is that the solution code submitted by a user should be evaluated in a coherent and reliable server infrastructure. Hence, these systems are developed as platform-as-a-service (PaaS) cloud computing services because the scalability of such a highly interactive system is crucial, especially as the competition deadline approaches and the number of submissions increases rapidly. For this reason, the efficiency of such systems is often guaranteed by an architecture utilizing concurrency and parallel processing. A detailed analysis of the utilization of symmetric multiprocessing (SMP) environments by online judges is presented by Drung et al. \citep{Drung_2011}.

Finally, not only is the software itself important, but so are the description of the problem and the quality of the prepared test cases stored in the system. Fori{\v{s}}ek noted that authors of problems, which are evaluated automatically during competitions such as the ACM ICPC and the IOI, should give special attention to both the type of the problem and the preparation of test cases. He demonstrated that some types of problems, such as substring search, can be easily solved using heuristic algorithms. Additionally, he presented several problems from the ICPC and IOI competitions that can be solved using generally incorrect algorithms and still receive very good evaluation score \citep{forivsek2006suitability}. Moreover, Mani et al. \citep{2014} noticed that the manner of presentation of the online judge output is also very important. In particular, the evaluation summary presented in online judge systems which educational institutions use during courses, should be easy to read and understand.

The problems that are usually published in online judge systems are generally classified as combinatorial problems. A combinatorial problem relies on discovering values of discrete variables that satisfy specific constraints. They are divided into decision problems, for which it must be verified that a provided solution exists, and search problems, for which the solution has to be found. A special case of the search problem is the optimization problem, for which we have to find the optimal solution subject to some objective function. Combinatorial problems are very interesting because they are intuitively understandable, but solving them is often challenging. The aforementioned observation was confirmed for various combinatorial problems and became a cause of the rise of computational complexity theory \citep{Edmonds1965,Garey1979}. This research field focuses on the classification of such problems, taking into account how complex the process of searching for a solution is. Problems published in online judge systems are usually solvable in polynomial time, and the maximal limit for processing time needed to find a solution is adjusted in such a way as to ensure that the optimal solution can be found for each considered test case. Although these systems can also be used to evaluate more complex problems for which only local optimums can be found during the processing time limit, none of the currently available online judges can simply manage such problems. The most well-known competition with the longest tradition is the ROADEF Challenge \citep{Artigues_2012} organized by the French Operational Research Society. The challenge is conducted every two years. Its primary objective relies on solving some industrial-driven problem proposed in collaboration with a business partner and supplemented with realistic test cases. There are also other specialized contests, usually focused on more specific topics, e.g., the International Planning Competition \citep{Coles_2012} organized during the International Conference on Automated Planning and Scheduling (ICAPS) focuses on the implementation of planners that solve instances of planning problems defined in the Planning Domain Definition Language (PDDL) format \citep{Mcdermott_1998}. Another example is the JILP Workshop on Computer Architecture Competitions, where participants solve problems related to optimizing parameters of algorithms used by computer hardware or infrastructure.

The aforementioned contests apply a very efficient technique called crowdsourcing to solve practical optimization problems. Crowdsourcing outsources work to a large network of people in the form of an open call \citep{Wasik_2015}. In the case of optimization challenges, the topic of the call is related to the optimization problem, and the network of potential participants includes interested programmers and scientists grouped together through the Internet. Although the concept of crowdsourcing has been intuitively applied since at least the eighteenth century \citep{Dawson_2012} it has been formally defined relatively recently in 2006 by Jeff Howe \citep{Howe_2006}. The first fifteen years of the twenty-first century have been a time of rapid development in crowdsourcing, thanks to the popularization of Internet access. The best example of success in solving industrial challenges using crowdsourcing is the Kaggle platform. It is a web portal where data mining problems are published and subsequently solved by external experts participating in the competition. In 2016, Kaggle completed 34 competitions with total awards valued at 1,160 thousand dollars. The number of teams participating in the considered competitions varied between 50 and 3,500. Assuming that the average number of teams participating in a single contest is equal to 500, the average size of a team is 3, and average time spent solving the problem is up to 3 days per person, thus yielding 135,000 days or 370 person-years spent solving industrial problems.

The Kaggle platform focuses mainly on data-mining problems related to the concept of data science, which has recently gained popularity \citep{Dhar_2013}. However, solutions of various optimization problems can also be practically applied, e.g., in problems originating from the operational research field. Thus, there is a great need to develop efficient algorithms to solve them in safe and reliable environments by a resilient community. In this article, we would like to review the state of the art in the field of online judge systems and briefly discuss the Optil.io platform as an example of such systems, designed with the application of crowdsourcing in mind to solve complex optimization problems. This implements the online judge concept that utilizes an objective function instead of a binary evaluation to rank submitted solutions. 

The scope of the article is as follows: As there is no formal definition of \textit{online judge} systems in literature, we, therefore, decided to fill this gap to clarify further analysis. Next, we present a survey of existing online judge systems, prepared with a focus on potential applications, and provide a short presentation of each of them. The diversity of these types of systems is so tremendous that the application-based classification is of great importance for interested practitioners. We focus mainly on systems that support an evaluation of algorithms for solving combinatorial problems. This is because, firstly, they are elementary type of problems from the theoretical point of view \citep{Garey1979}, secondly, they include many data mining problems \citep{Saedi2013} and, thirdly, they are most commonly addressed by this kind of systems. Additionally, we summarize the common evaluation methodology supported by these systems. Finally, we briefly demonstrate potential applications of the online judge systems based on the discussion of the contest conducted using the Optil.io platform. A brief discussion of the results of the application of this system allows for a broader understanding not only of the concepts described in this article, but also of their practical implications to the design and development of online judge systems. We believe that such a review, supplemented with an example application of an online judge system, can be interesting for many scientists who want to use online judge systems in their own research, or teachers who would like to use them for educational purposes. 

\section{Online judge systems}
\label{sec:state-of-the-art}

In general, the aim of online judge systems is a safe, reliable, and continuous, cloud-based evaluation of algorithms that are submitted by users distributed around the world. Toward a better understanding of the scope of this article, we will first define an evaluation procedure, which is crucial and should be implemented, at least partially, by any online judge.

\begin{definition}[Evaluation procedure]
\label{def:evaluation-procedure}
An evaluation procedure consists of three steps: (1) submission, (2) assessment, (3) scoring. 
\end{definition}

During the submission phase, the submitted code is compiled, if needed, and verified if it can be successfully executed in the homogeneous evaluation environment. After a successful verification, each submission is reliably assessed on coherent infrastructure based on the problem-specific set of test cases. For each test case execution of the particular submission it is verified whether:
\begin{enumerate}
\item the execution process proceeded without errors, 
\item any problem-specific resource limitations have not been exceeded,
\item the obtained output complies to the rules described in the problem definition.
\end{enumerate}
Finally, the aggregated score for the submission is computed based on the results of all considered test cases. The detailed definition of the commonly used evaluation procedure is presented in Section \ref{sec:evaluation-methodology}. We treat various online platforms as an online judge in a broad sense, including all systems that support any subset of evaluation procedure phases often in a cloud-based environment. A formal definition of online judge is provided in section \ref{sec:evaluation-methodology} in definition \ref{def:online-judge}.

\subsection{Methods}

In the literature, one can find various attempts at classification of contests organized using online judge systems and problems solved in such environments. In 2006, Pohl proposed a first simple classification \citep{Pohl_2006} taking into account criteria related to a contest's style, duration, grading, submission procedure, and entrance criteria. In 2014, Comb\selectlanguage{ngerman}é\selectlanguage{english}fis and Wautelet proposed another classification \citep{Combefis_2014} according to criteria focusing on the programming contests and the problems solved during these matches. In 2015, N\selectlanguage{ngerman}é\selectlanguage{english}meth et al. refined this classification with several new criteria, characterizing contests more deeply and describing types of programming exercises and features of online judge systems crucial to the educational perspective \citep{Nemeth_2015}. However, all these reviews had scope limited to a single application, either in education, or in the organization of programming contests. Until now, no one has classified online judge systems according to their potential applications, taking into account a much broader range of interests of potential users. Such classification should present the primary objective of the system and can be extremely useful for users looking for an online judge system that meets their needs. That is why we decided to treat this criterion as crucial to the differentiation of online judge systems. 

According to this criterion, we distinguished six classes of systems that integrate online judge systems. The largest class represents online platforms dedicated to the sharing of challenges solved during programming contests (ACM ICPC-like) and Olympiads in Informatics. The other categories of systems are helpful for educational purposes, recruitment of employees, evaluation of specialized algorithms that solve data mining problems and integration as a crucial component of online compilers that allow users to compile arbitrary code online. Finally, we present the class of development platforms that provide an online judge component that can be simply used in custom applications. In some cases, it was difficult to apply such a classification because there are several systems that can be considered as members of more than one group. In such cases, we identified the principal objective of the service to classify it. 

The objective of the article is to provide a tutorial demonstrating a broad range of online judge systems prepared based on an extensive review of literature conducted using the Web of Science, Google Scholar, and Scopus Indexes. Moreover, we reviewed in detail the vast majority of articles published in \emph{Olympiads in Informatics} journal and queried the Internet to find such systems that are available even if they have not yet been published. For all queries, we used the following set of keywords: \textit{online judge}, \textit{online judge systems}, \textit{automated programs grading}, and \textit{automated grading programming}. The area of students' assignments assessment has been quite extensively explored in many well-known review papers where such systems are described in the light of potential applications, e.g., teaching use cases \citep{caiza2013,Ihantola_2010,Romli2010,Staubitz2015,Ihantola2015,Wilcox2016}. To ensure completeness of the possible applications of online judges in the proposed review, we selected systems that are the most comprehensive, from the applications-oriented perspective, as representatives in this category.

Following the guidelines for preparing systematic
literature reviews \citep{kitchenham2009systematic,keele2007guidelines}, we defined the following acceptance criteria, all of which should be satisfied by the system considered in the review; i.e., the system:
\begin{enumerate}
\item supports any subset of evaluation procedure phases,
\item is able to evaluate algorithms that are used to solve combinatorial problems,
\item is available in English, or is described in an article published in English in some journal or conference proceedings,
\item is publicly available (for free or on commercial basis),
\item operates properly, i.e., it should have provided a possibility to register and submit solutions for at least one problem provided through it in June 2017.
\end{enumerate} 

A classification of online judge systems presented in this article has been prepared with the main focus on the summary of their possible applications from the user's perspective. For each online judge system that we found in the literature, we selected two to five of the most intuitive usage scenarios. Afterwards, for each usage scenario identified earlier, we computed a simple coverage coefficient as the number of online judge systems that implement such a scenario. Thus, we obtained a list of usage scenarios ordered in a decreasing manner according to the coverage coefficient. During the team brainstorming session, we discussed the significance and usability of the considered usage scenarios. Eventually, we defined the following categories of online judge systems that differ in supported functionalities:
\begin{enumerate}
\item Online compilers: systems that support only a compilation of source codes performed during a submission phase of the evaluation procedure.
\item Data mining, education and competitive programming: systems that, by inferring from our functionality-oriented analysis, focus on (a) evaluation of solutions submitted for data mining problems are classified in a data mining category; (b) educational processes are considered as educational software; (c) implementation of ICPC or IOI scoring rules and archiving problems from past challenges are classified as competitive programming software.
\item Recruitment platforms: online platforms targeted to employers to facilitate the recruitment of software developers for various IT projects, and which allow the evaluation and ranking of submitted solutions according to specific objective functions.  
\item Development platforms: systems that are often provided as open source projects or binary archives, allowing users to customize deployments and configurations of these systems in their infrastructure. 
\end{enumerate}

For each online judge system, we provide the most crucial information, namely if it is open source, what natural language it uses in its user interface, the number of supported compilers, number of published challenges, number of registered users, year of establishment, and if it is actively maintained. We included only systems that are available online and working correctly. We classified systems that had no new problems or news published on their websites over 2016 as inactive. Otherwise, a system is classified as active. When we could not locate information on the web page, we tried to contact administrators. The number of compilers supported by many platforms is changing very quickly because it is rather easy to extend the range of supported compilers. Thus, we assigned the considered systems into four classes, i.e., systems that do not support any compilers because they only assess the results obtained by algorithms executed by the user using his infrastructure (Class 0); systems that support a single specific programming language (Class 1); systems that support several of the most popular programming languages, such as C++, Java, Python, and C\# (Class 2); and systems that support a broad range of compilers (Class 3). For development platforms and online compilers, the assessment of the number of challenges published is not applicable because their business goal is to be integrated as part of more complex systems and to compile code that can be executed on arbitrary data. A summary of all this information is presented in Tables \ref{tbl:judges-competitive} to \ref{tbl:judges-development}.

\subsection{Competitive programming}

Online platforms that collect and share challenges similar to those used during competitive programming contests \citep{Khera_1993} constitute the largest group of services that integrate online judge systems because many universities provide this type of system to help their students prepare for competitive programming championships. Moreover, organizations that conduct such competitions are interested in the popularization of challenges solved during the past editions of these contests. The list of online judge systems presented in this subsection can be found in Table \ref{tbl:judges-competitive}.

{ \tiny
    \begin{longtable}{ cccccccc }
    
        \caption{{Online judge systems applied for competitive programming contests. Successive columns include: (1) the name and URL address available in electronic version when hovering the computer mouse over its name, for printing URL addresses please download the electronic supplement, (2) if this system is open source software (OSS), (3) the ISO 639-2/B code \citep{llc2010iso} of languages supported by the user interface of the system, (4) the class regarding the number of supported compilers, (5) the number of published problems, (6) the number of registered users, (7) the year of the establishment, and (8) the date of the last update of the system. All data were collected in June 2017.}}
        \label{tbl:judges-competitive} \\
        
        \hline \textbf{Name and URL address} & \textbf{OSS} & \textbf{GUI language} & \textbf{Compilers class id} & \textbf{\#Problems} & \textbf{\#Users} & \textbf{Established} & \textbf{Active} \\
        \hline 
        \href{https://a2oj.com/}{A2 Online Judge} & No & Eng & 3 & 300 & 55000 & 2011 & Yes \\
        \href{https://ac.2333.moe/}{AC 2333} & No & Chi & 2 & 670 & 3300 & 2011 & Yes \\
        \href{http://acdream.info/}{AcDream} & No & Eng,Chi & 2 & 300 & 5200 & 2013 & Yes \\
        \href{https://icpcarchive.ecs.baylor.edu}{ACM-ICPC live archive} & No & Eng & 3 & 1000 & 52000 & 2003 & Yes \\
        \href{http://www.olymp.krsu.edu.kg/GeneralProblemset.aspx}{ACM-Kyrgyzstan Subregion} & No & Eng & 3 & 422 & 3600 & 2005 & Yes \\
        \href{https://adjule.pl/}{Adjule Online Judge} & No & Pol & 2 & 120 & 3000 & 2011 & Yes \\
        \href{http://judge.u-aizu.ac.jp/onlinejudge/}{Aizu Online Judge} & No & Eng,Jpn & 3 & 1000 & 36000 & 2004 & Yes \\
        \href{http://www.azspcs.net/}{Al Zimmermann's Programming Contests} & No & Eng & 0 & 26 & 2000 & 2009 & Yes \\
        \href{http://www.bnuoj.com/v3/index.php}{BNU OJ} & Yes & Eng,Chi & 3 & 51000 & 31000 & 2013 & Yes \\
        \href{http://coj.uci.cu/index.xhtml}{Carribean Online Judge} & No & Eng,Spa & 3 & 2700 & 28000 & 2010 & Yes \\
        \href{http://acm.uestc.edu.cn}{CDOJ} & Yes & Eng,Chi & 2 & 1300 & 9600 & 2014 & Yes \\
        \href{http://codeforces.com/}{Codeforces} & No & Eng,Rus & 3 & 3000 & 32500 & 2010 & Yes \\
        \href{https://dmoj.ca/problems/}{Don Mills Online Judge} & Yes & Eng & 3 & 700 & 7700 & 2014 & Yes \\
        \href{http://e-olymp.com}{e-olymp} & No & Eng & 3 & 7500 & 47000 & 2006 & Yes \\
        \href{http://acm.mipt.ru/judge}{EI Judge} & No & Eng,Rus & 3 & 400 & 20000 & 2003 & Yes \\
        \href{https://www.facebook.com/hackercup}{Facebook Hacker Cup} & No & Eng & 0 & N/A & 80000 & 2011 & Yes \\
        \href{http://acm.fzu.edu.cn/}{Fuzhou University Online Judge} & No & Eng,Chi & 2 & 1300 & 34000 & 2008 & Yes \\
        \href{https://code.google.com/codejam/}{Google Code Jam} & No & Eng & 0 & 450 & 200000 & 2008 & Yes \\
        \href{http://herbert.tealang.info/problems.php}{Herbert Online Judge} & No & Eng,Chi & 1 & 1761 & 1200 & 2010 & 2011 \\
        \href{http://acm.hit.edu.cn/hoj}{HIT ACM/ICPC} & No & Eng & 2 & 1300 & 54000 & 1998 & Yes \\
        \href{http://acm.hust.edu.cn/}{HUSTOJ} & Yes & Eng,Chi & 2 & 650 & 26000 & 2014 & Yes \\
        \href{http://opc.iarcs.org.in/index.php/problems/}{Indian Computing Olympiad Problems Archive} & No & Eng & 2 & 45 & 1250 & 2003 & Yes \\
        \href{http://ipsc.ksp.sk}{Internet Problem Solving Contest} & No & Eng & 0 & 240 & 5000 & 1999 & Yes \\
        \href{http://lightoj.com}{Light OJ} & No & Eng & 2 & 430 & 14000 & 2012 & Yes \\
        \href{http://www.lydsy.com/JudgeOnline/}{LYDSY} & No & Chi & 2 & 1000 & 30000 & 2008 & Yes \\
        \href{http://main.edu.pl}{Main} & No & Eng,Pol & 1 & 1000 & 33000 & 2005 & Yes \\
        \href{http://acm.csie.ntu.edu.tw/ntujudge/index.php}{National Taiwan University Online Judge} & No & Chi & 2 & 2600 & 600 & 2016 & Yes \\
        \href{http://acm.cs.nthu.edu.tw/}{National Tsing Hua University Online Judge} & No & Eng & 1 & 10000 & - & 2015 & Yes \\
        \href{https://noj.io/}{North University of China Online Judge} & No & Eng & 1 & 2000 & 4000 & 2006 & Yes \\
        \href{http://wcipeg.com/}{P3G} & No & Eng & 3 & 1100 & 500 & 2008 & Yes \\
        \href{http://poj.org/}{Peking University Judge Online} & No & Eng & 2 & 3000 & 250000 & 2003 & Yes \\
        \href{http://acm.petrsu.ru}{Petrozavodsk State University} & No & Eng,Rus & 2 & 450 & 140 & 2010 & Yes \\
        \href{https://projecteuler.net/}{Project Euler} & No & Eng & 0 & 550 & 650000 & 2001 & Yes \\
        \href{http://www.spoj.com/}{SPOJ} & No & Eng & 3 & 6000 & 60000 & 2004 & Yes \\
        \href{http://pl.spoj.com/}{SPOJ PL} & No & Pol & 3 & 800 & 36000 & 2004 & Yes \\
        \href{http://szkopul.edu.pl/}{Szkopuł} & Yes & Pol & 1 & 1000 & - & 2012 & Yes \\
        \href{https://www.teddyonlinejudge.net/}{Teddy Online Judge} & Yes & Spa & 3 & 250 & 1900 & 2009 & Yes \\
        \href{http://acm.timus.ru/}{Timus Online Judge} & No & Eng & 3 & 1000 & 110000 & 2000 & Yes \\
        \href{http://acm.tju.edu.cn/toj/}{TJU ACM-ICPC Online Judge} & No & Eng,Chi & 2 & 3000 & 52000 & 2005 & Yes \\
        \href{https://www.topcoder.com/community/competitive\%20programming/}{TopCoder Competitive Programming} & No & Eng & 2 & 5200 & 4000 & 2001 & Yes \\
        \href{http://usaco.org}{USA Computing Olympiad} & No & Eng & 2 & 150 & 12000 & 2013 & Yes \\
        \href{http://uva.onlinejudge.org/}{UVa Online Judge} & No & Eng & 2 & 5000 & 250000 & 1995 & Yes \\
        \hline
    \end{longtable}
}

The first online judge system that gained high popularity worldwide is the \oj{UVa Online Judge} \citep{Revilla_2008}. It was founded in 1995 by Miguel  \selectlanguage{ngerman}Á\selectlanguage{english}ngel Revilla, a mathematician who lectures on algorithms at the University of Valladolid in Spain. It provides an enormous archive of programming challenges originating from the ACM contests. Inspired by the massive data set collected by UVa, Skiena and Revilla wrote their first book, the objective of which was primarily to help students prepare for team programming competitions \citep{Skiena_2008}. In this volume, they presented a large subset of challenges, followed by their most compelling solutions and hints that can be used by readers to solve them themselves.

Currently, the most extensive online judge, where over 10,000 challenges have been published, is the \oj{National Tsing Hua University Online Judge}. It is an ACM ICPC-like online judge designed for training purposes as well as the online platform often used for organization of programming contests. It supports only the C and C++ programming languages. In turn, one of the most popular online judges is \oj{Codeforces}. It organizes contests on a regular basis. Moreover, it provides independent instances prepared for Russians only. It divides participants into two divisions according to their adaptation of Elo rating \citep{elo1978rating} which is updated for every competition conducted. During the contest, they randomly split contestants into rooms of approximately 40 people. A single round takes 2 hours and considers five programming challenges. After final submission of the solution for a given challenge, it allows each contestant to inspect other participants' solutions to \textit{hack} them. By \textit{hack}, we mean that Codeforces allows the participants to verify the code submitted by rivals on an instance of their choice. When the verification fails, the system automatically rewards the contestant who demonstrated the failure of the submitted solution with additional points, and the author of the \textit{hacked} one receives a penalty. 

Three other large online platforms with over 5,000 published problems are \oj{Sphere Online Judge} (\oj{SPOJ}), \oj{E-Olymp} and \oj{TopCoder Competitive Programming}. The first \citep{Kosowski_2007} is a very popular system that offers various types of challenges, ranging from classical to optimization, code-golf, and riddles. It also provides two independent instances of the system dedicated for Polish and Brazilian users. Each of these instances includes separate challenges. In turn, E-Olymp is a Ukrainian portal supporting national institutions responsible for teaching gifted young people who participate in programming competitions worldwide. Finally, TopCoder provides a very innovative way of conducting algorithmic contests. They organize short rounds, with three challenges and a 75-minute time limit per round. For each submitted solution that proves to be correct, the authors receive a reward inversely proportional to the time that they needed to solve it and an additional bonus when they identify an issue in an opponent's code, similarly to Codeforces. TopCoder is a service with a very long tradition and has already organized over a thousand algorithmic matches.

There are also plenty of less-popular online judges maintained by various universities. For example, \oj{Tianjin University's Online Judge} (\oj{TJU}) provides an extensive database of challenges, which can be used to conduct contests that are custom, open or addressed to a limited set of participants free of charge. These are so-called virtual contests. The organizer of such a contest is responsible for choosing its start and end date and selecting the set of challenges from the provided database that should be solved during this contest. Next, the system hosts the contest, following the rules defined for the ACM ICPC contests. In this context, it is also worth mentioning \oj{Peking University} (\oj{PKU}) \oj{Judge Online} (\oj{POJ}) and \oj{Timus Online Judge}, which are the largest online judge systems located in China \citep{Wen_2005} and Russia, respectively. \oj{EI Judge} is a relatively small system of that type in comparison with predecessors hosted by the Moscow Institute of Physics and Technology. In turn, \oj{AC 2333} is the Ningbo University of Technology's online judge system, which only operates in China. The first Polish online judge in this grid is the \oj{Adjule} system hosted by Adam Mickiewicz University, which mainly shares challenges proposed for the annual contest called the Poznan Open Championships. The following systems, namely those of Petrozavodsk State University (which hosts the \oj{PetrSU Programming Club}) and \oj{National Taiwan University}, are used to organize programming contests for their students. In contrast to the systems mentioned above, \oj{P3G} is a resource for competitive programmers worldwide and was created by members of Woburn Collegiate Institute's Programming Enrichment Group. Creators of this platform also run \oj{PEGWiki}, which is a site prepared for algorithm enthusiasts. Other universities that boast their own online judges include \oj{Fuzhou University} (China), \oj{Harbin University of Technology} (China), \oj{University of Electronic Science and Technology of China}, \oj{North University of China}, University of Information Science in Cuba (hosting \oj{Caribbean Online Judge}), and Huazhong University of Science and Technology in China (hosting \oj{HUSTOJ}).

Moreover, many organizations that host programming competitions are also interested in managing online judge systems to popularize challenges solved in the past events and help users prepare for upcoming ones. One such system is the USA Computing Olympiad (\oj{USACO}) platform, where a vast number of programming challenges and well-written tutorials covering various examples are published \citep{kolstad2007usa}. It introduces the world of competitive programming step by step. Moreover, it organizes five or six online contests each year. The \oj{ACM-ICPC Live Archive} is an ACM programming challenges database originating from regional contests, as well as finals held worldwide since 1988. Any interested participant can solve those problems in numerous programming languages. In turn, the users of the \oj{Aizu Online Judge} solve challenges originating from both the Japanese Olympiad in Informatics and Japanese high school contests. The \oj{MAIN} website provides an archive of various challenges originating from Polish and international contests co-organized by the Polish Olympiad in Informatics. Moreover, it also offers interactive courses on programming and algorithms. A platform supplementing MAIN is \oj{Szkopul}, which allows users to create virtual contests using challenge sets hosted on MAIN. There are also other, smaller, systems, such as the \oj{ACM Kyrgyzstan Subregion Challenges Archive}, \oj{Indian Programming Olympiad Archive}, \oj{A2 Online Judge}, \oj{AcDream}, and \oj{Light OJ}.

Unfortunately, most of the systems mentioned above are closed source, except HUSTOJ. There are also three other platforms that are open source: \oj{CDOJ}, \oj{Teddy Online Judge}, and \oj{Don Mills Online Judge (DMOJ)}. The last one especially is worth mentioning in this category because it is a fully open-source and well-documented system available on GitHub. It publishes challenges originating from Canadian Computing Competition, Canadian Computing Olympiad, Croatian Open Competition in Informatics, International Olympiad in Informatics, and many others. It also runs a contest once every month. Thanks to the extensive documentation, it is also very easy for potential newcomers to submit novel challenges to DMOJ.

There are also several websites that present less classical applications of online judge systems, e.g., \oj{Project Euler}, which hosts more specific mathematical challenges in comparison with regular ACM counterparts. Every week, a new challenge is posted. Users are ranked according to their quality scores obtained for submitted solutions. The quality score depends on the number of users who have submitted the correct solution earlier. Global ranking is constructed by taking into account the quality scores obtained for the ten most recent challenges. Users usually develop a dedicated program. However, in general, they can even try to solve the current challenge analytically because only the textual representation of the solution obtained for the particular test instance can be submitted and automatically judged. In turn, \oj{Al Zimmermann's Programming Contests} is another online platform where contestants are supposed to send only the textual representation of the solution for computationally intensive problems. It runs new contests once or twice a year. Another event worth mentioning is the \oj{Internet Problem Solving Contest} (\oj{IPSC}), which is an annual programming competition considering fancy challenges of various types, e.g., the ACM, Capture the Flag, optimization problems, etc. Contestants are grouped together in teams of three. Those who sent a postcard to coordinators are granted an additional time bonus. At the end of a particular contest, all challenges are moved to the archive where they are available for practicing as part of a virtual contest system. Another interesting system is the \oj{Herbert Online Judge}, which provides over a thousand challenges that are solved by users in the H language. The H language is a very simple programming language proposed for controlling a robot called Herbert that was developed for the purpose of the challenge called the Algorithm Competition conducted during the Imagine Cup tournament in 2008. It allows the participant to strengthen his/her algorithmic skills, especially regarding the finding of patterns and implementing of recursion. Finally, two of the most prominent IT companies, Facebook and Google, also organize their own programming competitions. \oj{Facebook Hacker Cup} is an annual competition that uses an online judge system to evaluate the quality of contestants' submissions. Unfortunately, the challenge archive is not available publicly when the particular contest ends. \oj{Google Code Jam} is an event that is also organized annually. Sometimes there are also regional or even special editions. During the competition, participants send the textual representation of the solution computed for the particular test instance and the program sources that were used to obtain it. However, the latter are used only for verification if participants do not share solutions among each other, and only the former is evaluated.

Because there are thousands of challenges collected in numerous online judge systems provided on the Internet, there have also been some attempts to index and classify them. Zhu and Fu \citep{Zhu_2012} introduced a system for automatic classification of challenges based on hierarchical knowledge representation \citep{Yoon_2006}. Unfortunately, their system has not been put into practice. There are also other online resources maintained manually that could be helpful in choosing interesting challenges from among the many provided examples, such as \oj{uHunt} dedicated for the UVA online judge. The very useful initiative emerged in the \oj{Beijing Normal University Online Judge} (\oj{BNUOJ}). It provides a single local judge equipped with slightly over 1000 challenges and a virtual one that allows the user to apply 24 other online judges. In this manner, the system integrates over 50,000 challenges published on various websites. When the user submits a solution to one of these remotely available challenges, the BNUOJ automatically forwards it to the appropriate online judge and acquires from the remote system the evaluation result. It also supports the inclusion of the challenge phase similarly to the Codeforces and TopCoder Competitive Programming contests. Moreover, it allows users to organize virtual competitions using challenges originating from local as well as remote archives. This system can replay contests and merge the final standings. Last but not least, the BNUOJ is an open-source project available on GitHub.

\subsection{Education}

The application of online judge systems in education has a very long history, as they have been used at least since 1961 when they were introduced at Stanford University \citep{Leal98automaticgrading,Forsythe_1965} to support the evaluation of students' programs coded in ALGOL. This concept was followed by many more systems, no existing currently, such as Ceilidh \citep{Benford_1993}, Kassandra \citep{von_Matt_1994}, CoBalT \citep{Joy_1995} or RoboProf \citep{Daly_1999} and finally resulted in the implementation of online judges in the form of a massive open online course (MOOC). In general, a MOOC is addressed to an unlimited number of participants that can learn independently worldwide without the requirement of respecting a fixed course schedule \citep{Pieterse_2013}. We can distinguish two main types of MOOCs, namely traditional online courses and those requiring collaboration between participants and teaching staff, i.e., xMOOC \citep{Prpic_2015} and cMOOC \citep{Kop_2011}, respectively. The former usually provides a fixed set of recorded lectures and self-test exercises. In the case of online judge systems, these exercises are presented in the form of automatically judged programming tasks. The latter is much more successful in knowledge building as a consequence of a collaborative dialogue \citep{Bell_2011}. In the case of online judge systems, such dialogue is usually implemented by the possibility of adding annotations to the students' source code. In 2012, several of the most well-known organizations in this field emerged, i.e., edX (a non-profit organization formed by MIT and Harvard), Coursera and Udacity (originating from Stanford). In the middle of 2013, edX and Google started a partnership to establish a commonly used resource (i.e., mooc.org) dedicated to building and hosting various e-learning courses. Currently, the numbers of courses provided by the organizations mentioned above are on the order of hundreds or even thousands. Development of MOOC is not cheap for organizations but, in the long term, yields savings. Moreover, it allows them to differentiate from other educational organizations, even without an appropriate level of expertise. However, it is worth noticing that MOOC requires a high degree of motivation and self-discipline from the participants. Thus, it is usually a way to strengthen professional skills, and it is rarely used to learn a new field from scratch \citep{Kaplan_2016}.  

Utilization of online judges allows an educational organization's staff to assess students' assignments automatically. Application of these systems results in several advantages. First, the teacher can verify the correctness of solutions submitted by students with higher accuracy. When the teacher prepares the complete set of test instances covering all corner cases resulting from the problem definition, the possibility of acceptance of an incorrect solution is almost negligible. Second, the time needed for evaluation is much shorter; therefore, the teacher can prepare and assign to students many more exercises. Finally, students receive an almost instant answer as to whether their solution is correct. An inspiring description of the successful application of an online judge in the teaching of Algorithms and Data Structures and Competitive Programming courses at the National University of Singapore is presented in \citep{Cheang_2003}. Ala-Mutka presented an in-depth review of applications of online judges in education based on the analysis of several systems \citep{Ala_Mutka_2005}. In 2010, Ihantola et al. prepared a more recent review of available software dedicated to automatic assignment of programming exercises, focusing on detailed descriptions of their features and various usage scenarios, which is interesting from a pedagogical or educational point of view \citep{Ihantola_2010}. In turn, Cruz et al. presented how to extend the typical online judge architecture to develop a system that goes a step further by providing valuable feedback to users at a semantic level in the form of meaningful advice to understand where the problem is and how to improve the code \citep{Daniela_2013}. Such an approach can significantly impact the student learning process \citep{Wang_2016}. Finally, there is an interesting idea of building computer games that integrate or wrap online judge systems to be more attractive to users \citep{Ivanova_2016}. The list of existing online judge systems presented in this subsection can be found in Table \ref{tbl:judges-education}.

{ \tiny
    \begin{longtable}{ cccccccc }
    
        \caption{{Online judge systems applied for educational purposes. For column descriptions, see Table \ref{tbl:judges-competitive}.}}
        \label{tbl:judges-education} \\
        
        \hline \textbf{Name and URL address} & \textbf{OSS} & \textbf{GUI language} & \textbf{Compilers class id} & \textbf{\#Problems} & \textbf{\#Users} & \textbf{Established} & \textbf{Active} \\
        \hline 
        \href{http://www.checkio.org}{CheckiO} & No & Eng & 1 & 100 & 110000 & 2013 & Yes \\
        \href{https://codefights.com/}{Code Fights} & No & Eng & 3 & 1250 & 500000 & 2015 & Yes \\
        \href{https://codeboard.io/}{Codeboard} & Yes & Eng & 3 & 24000 & 60000 & 2015 & Yes \\
        \href{https://www.codecademy.com/}{Codecademy} & No & Eng & 3 & N/A & 25000000 & 2011 & Yes \\
        \href{http://www.codechef.com/}{CodeChef} & No & Eng & 3 & 1500 & 300000 & 2009 & Yes \\
        \href{http://www.codehunt.com}{CodeHunt} & No & Eng & 1 & 134 & 350000 & 2014 & Yes \\
        \href{http://www.codewars.com}{Codewars} & No & Eng & 3 & 1200 & 400000 & 2012 & Yes \\
        \href{http://www.codingame.com/start}{CodinGame} & No & Eng & 3 & 55 & 500000 & 2012 & Yes \\
        \href{http://codingbat.com/}{CodingBat} & No & Eng & 2 & 300 & - & 2009 & Yes \\
        \href{https://microcorruption.com}{Embedded Security CTF} & No & Eng & 1 & 19 & 35000 & 2014 & Yes \\
        \href{http://exercism.io}{Exercism} & Yes & Eng & 3 & 1450 & 30000 & 2013 & Yes \\
        \href{https://jutge.org/}{Jutge.org} & No & Eng,Spa,Cat,Ger,Fre & 3 & 2000 & 14000 & 2006 & Yes \\
        \href{https://leekwars.com/}{Leek Wars} & No & Eng,Fre & 1 & 1 & 54000 & 2013 & Yes \\
        \href{http://programming.grids.cn/programming/}{Programming Grid} & No & Chi & 2 & 640 & - & 2008 & Yes \\
        \href{http://www.pythonchallenge.com/}{Python Challenge} & No & Eng & 0 & 33 & N/A & 2005 & Yes \\
        \href{https://racso.lsi.upc.edu/juez/}{RACSO} & No & Eng,Spa,Cat & 0 & 330 & - & 2012 & 2015 \\
        \href{http://theaigames.com/}{The AI Games} & No & Eng & 3 & 8 & 2700 & 2013 & Yes \\
        \href{https://www.urionlinejudge.com.br/judge/login}{URI Online Judge} & No & Eng,Spa,Por & 2 & 1170 & 4100 & 2011 & Yes \\
        \hline
    \end{longtable}
}

Because online judges that publish challenges collected from competitive programming contests have become very popular these days, there also exist sophisticated adaptations of such systems for educational purposes. For example, the \oj{URI Online Judge} is a system where provided challenges are distinguished into eight categories, allowing users to easily find exercises from a given topic. It also provides a panel addressed for teaching staff that supports tracing of the student learning process. Another adaptation worth mentioning is \oj{CodeChef}. It is a system that provides an online code editor and compilers that support over 40 programming languages. It classifies all provided challenges into categories according to their difficulty levels. However, even the easiest ones are relatively difficult. Much easier tasks can be found at \oj{Jutge.org} \citep{Petit_2012}, which is an online judge (free for educational purposes) where students can solve thousands of challenges using approximately 20 different programming languages, including even less popular ones, such as Verilog. It provides sophisticated panels for students and instructors. Finally, \oj{Programming Grid} is an interesting online platform that allows users to organize educational courses based on a challenges archive provided by the Peking University Judge Online. It follows a course-based concept instead of a challenge-based concept \citep{Luo_2008}. Unfortunately, this service is only available in Chinese.

Some of the online judges introduce additional gamification elements \citep{Deterding_2011} to strengthen the commitment and excitation of the students. For example, \oj{CheckiO} is a platform that provides a large number of relatively easy tasks prepared to support the learning process of the Python and JavaScript programming languages. It provides a few elements of gamification and social networking. Moreover, it allows the users to define ``classrooms'' and assign particular students to them to support the process of monitoring their learning progress. The platform was designed in a very attractive way and is still actively extended with new features. CheckiO is also supplemented with a massively multiplayer online strategy game called \oj{Empire of Code} where players can solve various types of problems collected by the CheckiO platform to gain various in-game bonuses. To be successful, they also have to develop artificial bots that are responsible for the protection of their units, as well as attacking their potential enemies. Another platform in this category that is worth mentioning is \oj{Codewars}. It collects a large number of exercises ordered by both difficulty level and topic-oriented categories. Codewars always provides several basic test cases, which are used for preliminary verification of the solution. A unique feature of this platform is that its user interface is inspired by Japanese culture (e.g., kata, kyu, and Kumite). Finally, \oj{CodeHunt} is a game developed by Microsoft to strengthen the algorithmic skills of players. It provides a snippet of code together with corresponding test cases. The goal of the player is to implement, in C\# or Java, an algorithm that will generate the expected output for all provided test cases.

There are also two additional educational games worthy of consideration in this category. In \oj{Leek Wars}, the player is responsible for implementation of an artificial intelligence script in LeekScript, a specialized programming language designed especially for Leek Wars, to win in a turn-based game. By defeating the opponents in the game, the user can upgrade his personal features, which allow him to be significantly better armed. It is also possible to improve the AI script based on the experience gained from skirmishes against provided bots, as well as AI scripts implemented by other users. In turn, \oj{PythonChallenge} is another in-browser game where the user develops the solution to simple riddles following hints provided in real-time by the system.

There are also other online platforms supporting less classical attitudes for the evaluation of submitted solutions to programming exercises. \oj{Codecademy} is one of them, focusing mainly on providing programming courses. Instead of exercises, the platform provides many professional tutorials. To progress in the given course, the user has to finish all the steps considered by the particular tutorial. The platform provides a code editor, a terminal, and other practical tools required to practice newly learned skills. In turn, \oj{CodinGame} is an online platform providing a variety of programming challenges. It supports programming puzzles, optimization and code-golf problems, and multiplayer games where the user implements an artificial intelligence script to control the bot. The significant advantage of this website over the counterparts is the vast number of supported programming languages and the visually attractive and exciting challenges, guaranteeing much fun and satisfaction during the solving of them. It also continuously gives the opportunity to participate in so-called code clashes, which represent very short competitions organized for several users that are started automatically by the system many times a day. Another platform, \oj{CodingBat} is a very simple page for people who have just started programming. This site supports only the Python and Java programming languages. It is not even necessary to be registered to submit solutions. Another tool called \oj{Exercism.io} provides a very original approach for the solving of programming assignments. It provides a command line tool and an API that is used to fetch task descriptions and submit solutions. The problem definitions are prepared in such a way as to follow the Test Driven Development (TDD) methodology. Authors of this platform provide only a general description of the problem and the automated test suite. The task of the user is to deduce all relevant details from the inspected test suite. Finally, \oj{Codeboard} is a user-friendly system that supports, not only the teaching of various programming languages by creating and sharing exercises with students, but also a test case based automated evaluation that can be easily customized by a teacher. Moreover, the teacher can easily follow the students' progress \citep{Antonucci_2015}.

One of the unique features of the CodinGame platform described above is the opportunity to compete in multiplayer games. In such games, the users develop artificial intelligence scripts that control bots competing with other bots in the virtual arena generated by the system. There are also two additional portals that follow such a concept, namely \oj{Code Fights} and \oj{AI Games}. The former is a platform where users develop their AI scripts in one of nine modern programming languages. Users receive experience points for every victory of their bots in duels. Based on gained points, users rise in level and earn badges. The idea behind the latter platform is to create AI scripts for playing most popular games, such as Tic-Tac-Toe or Go. This platform provides eight popular game engines and judges. Users can submit an AI script representing the particular bot that will compete in the given game with AI scripts provided by other users. Users are highly motivated to continuously improve their AI scripts to climb in the rankings. In fact, the system is closed source. However, the source code of engines is shared on GitHub.

Finally, two systems support the strengthening of professional skills in highly sophisticated programming concepts. The objective of the \oj{Embedded Security CTF} system represents support for learning the assembly language during a password cracking game. The user has to discover the password by analyzing the assembly program using the provided debugger to inspect the details of the password verification process. Based on the performed analysis, the user has to guess the password that will be accepted by the system. The tools provided by the platform are advanced, e.g., the user can analyze the entire code during its execution, set breakpoints and inspect values stored in processor registries and memory dumps. In turn, \oj{RACSO} is a platform where exercises related to automata and formal languages are provided \citep{Creus_2014}. In this system, the judge accepts the description of the automaton, as well as grammar.

\subsection{Online compilers}

Another category of online judges defined in a broad sense according to definition \ref{def:online-judge}, represents online platforms where user source code, developed in various programming languages, can be remotely compiled and executed via browser. In fact, they do not allow the publishing of any challenges because they usually support only the first step of the evaluation procedure provided in the definition \ref{def:evaluation-procedure}. Sometimes, partial support is also included for the assessment phase by allowing the user to submit his own test instances, which are used during evaluation. The list of online compilers discussed in this subsection is presented in Table \ref{tbl:judges-online-compilers}.

{ \tiny
    \begin{longtable}{ cccccc }
    
        \caption{{Online compilers. For column descriptions, see Table \ref{tbl:judges-competitive}. Online compilers do not provide a problem database and usually do not require registration, thus, the columns related to the number of problems and users have been removed.}}
        \label{tbl:judges-online-compilers} \\
        
        \hline \textbf{Name and URL address} & \textbf{OSS} & \textbf{GUI language} & \textbf{Compilers class id} & \textbf{Established} & \textbf{Active} \\
        \hline 
        \href{http://cpp.sh/}{C++ Shell} & No & Eng & 1 & 2014 & Yes \\
        \href{https://codeanywhere.com}{Codeanywhere} & No & Eng & 3 & 2013 & Yes \\
        \href{http://codepad.org/}{Codepad} & No & Eng & 3 & 2008 & Yes \\
        \href{http://www.codeskulptor.org/}{CodeSkulptor} & Yes & Eng & 1 & 2012 & Yes \\
        \href{http://www.tutorialspoint.com/codingground.htm}{Coding Ground} & No & Eng & 3 & 2006 & Yes \\
        \href{http://codio.com}{Codio} & No & Eng & 3 & 2013 & Yes \\
        \href{https://ideone.com/}{Ideone} & No & Eng & 3 & 2009 & Yes \\
        \href{http://www.onlinecompiler.net/}{Online Compiler} & No & Eng & 2 & 2009 & 2013 \\
        \href{http://webcompiler.cloudapp.net/}{Web Compiler} & No & Eng & 1 & 2014 & Yes \\
        \hline
    \end{longtable}
}

One of the most feature-rich online compilers is \oj{Codeanywhere}, which is a cloud-based IDE that allows users to share a dedicated virtual development environment and collaborate in real-time. It provides the ability to connect automatically to GitHub, Bitbucket, FTP server and Amazon cloud. A user can also set up a single, specialized, virtual container where a custom development environment will be created out of the box for free. This environment is very stable, even when many developers collaborate within the same project in real-time. Another online platform offering a fully featured IDE is \oj{Coding Ground}. It allows users to edit, compile, execute and share their projects in a cloud-based environment. It provides free terminals and IDEs that support development in plenty of different programming languages. Every program is executed in a dedicated Docker-based container created on demand. Finally, \oj{Codio} is a cloud-based online IDE supporting a large number of programming languages. It provides many useful features, such as remote Ubuntu-based development machine, integration with e-learning platforms, and plagiarism detection.

There are also several simpler online compilers, for which the objective is to provide the opportunity to compile and verify user code quickly without the need to construct separate virtual containers for specialized purposes. For example, \oj{Ideone} is a freely accessible online compiler that supports plenty of programming languages and is maintained by the authors of the SPOJ online judge. In turn, \oj{CodeSkulptor} is an online interpreter of the Python programming language that has a very aesthetic user interface and supports the learning process, especially for beginners. Moreover, it visualizes the execution of the particular program and provides an unofficial open-source offline server that can be used to run CodeSkulptor locally. Another system in this category called \oj{C++ Shell} provides an online interface for the GCC compiler. This system allows the user to compile the submitted source code and run it in a virtual sandbox environment created on demand. \oj{Codepad}, in comparison to the other counterparts, allows users to share code among collaborators using custom URLs. Another system worth mentioning is called \oj{Online Compiler}, which is a platform that allows users to remotely compile submitted source code developed in C/C++, Fortran, Java, Pascal or Basic programming languages and download the executable file built for Windows or Linux. Finally, \oj{Web Compiler} is an online compiler for Visual C++ that provides a minimalistic user interface.

\subsection{Recruitment}

There are also several, mainly commercial, platforms that use online judge systems primarily to support the recruitment process. The list of systems discussed in this subsection is presented in Table \ref{tbl:judges-recrutiment}.

{ \tiny
    \begin{longtable}{ cccccccc }
    
        \caption{{Online judge systems applied for recruitment purposes. For column descriptions, see Table \ref{tbl:judges-competitive}.}}
        \label{tbl:judges-recrutiment} \\
        
        \hline \textbf{Name and URL address} & \textbf{OSS} & \textbf{GUI language} & \textbf{Compilers class id} & \textbf{\#Problems} & \textbf{\#Users} & \textbf{Established} & \textbf{Active} \\
        \hline 
        \href{https://www.codeeval.com/}{CodeEval} & No & Eng & 3 & 240 & 85000 & 2009 & Yes \\
        \href{https://codility.com}{Codility} & No & Eng & 3 & 3000 & - & 2009 & Yes \\
        \href{https://www.hackerearth.com/}{HackerEarth} & No & Eng & 3 & 3700 & 1000 & 2012 & Yes \\
        \href{https://www.hackerrank.com}{Hackerrank} & No & Eng & 3 & 1000 & 84000 & 2009 & Yes \\
        \href{http://leetcode.com/}{LeetCode Online Judge} & No & Eng & 3 & 190 & - & 2010 & Yes \\
        \href{https://qualified.io/}{Qualified} & No & Eng & 3 & 4500 & - & 2015 & Yes \\
        \hline
    \end{longtable}
}

First, \oj{CodeEval} is a platform used by developers to showcase their skills in application-building competitions and programming challenges. CodeEval represents an exclusive community of developers who can compete and, as a result, build out their profiles to showcase their coding skills in the software development community. It uses Docker-based containers constructed on demand. Another platform worth mentioning is \oj{Codility}, which supports recruiters in reaching out to a large number of promising candidates in a relatively short amount of time. Moreover, developers can strengthen their coding skills by competing in programming challenges to build their professional reputation within the community. In turn, \oj{HackerEarth} is an online platform for which the main objective is hiring talented developers, organizing hackathons, and hosting crowdsourcing-based ideas. Developers can win various types of rewards. From companies' point of view, it is an excellent way to gather innovative ideas from a diverse community, including developers as well as computer scientists. \oj{HackerRank} is an online tool similar to the former one and was designed with supporting the hiring of developers in mind. It focuses mainly on supporting recruiters with the delivery of fully customized coding challenges addressed directly to the potential candidates and the seamless integration of real-time assessments into the recruiting process. From developers' point of view, it is a place where people from all over the world can solve rather tricky programming challenges classified into several categories, namely algorithms, machine learning and artificial intelligence. Finally, \oj{Qualified} is a platform provided by the authors of Codewars. Here, the developer implements the solution in real-time, and then the system automatically executes and qualifies it based on a predefined unit test cases set. It provides an interactive IDE, which supports an incremental development. During real-time interviews, the recruiter can benefit from the application of whiteboard tests. On the other side, there is another very useful platform called \oj{LeetCode}, which is a web service that supports preparation for technical job interviews. Users can solve exercises that may occur during such interviews divided into the following categories: algorithms, databases, and Linux shell.

\subsection{Data-mining services}
\label{ssec:data-mining}

There are also various websites that uses online judge systems to evaluate data-mining algorithms. Their list is presented in Table \ref{tbl:judges-data-mining}.

{ \tiny
    \begin{longtable}{ cccccccc }
    
        \caption{{Online judge systems applied for data-mining purposes. For column descriptions, see Table \ref{tbl:judges-competitive}.}}
        \label{tbl:judges-data-mining} \\
        
        \hline \textbf{Name and URL address} & \textbf{OSS} & \textbf{GUI language} & \textbf{Compilers class id} & \textbf{\#Problems} & \textbf{\#Users} & \textbf{Established} & \textbf{Active} \\
        \hline 
        \href{https://www.crowdanalytix.com/}{CrowdANALYTIX} & No & Eng & 0 & 105 & 16000 & 2012 & Yes \\
        \href{http://dreamchallenges.org/}{DREAM Challenges} & No & Eng & 0 & 45 & 5000 & 2006 & Yes \\
        \href{http://www.kaggle.com}{Kaggle} & No & Eng & 0 & 220 & 550000 & 2010 & Yes \\
        \href{http://mlcomp.org/}{MLcomp} & No & Eng & 1 & 12400 & 7000 & 2010 & 2017 \\
        \href{http://www.openml.org/}{OpenML} & Yes & Eng & 3 & 19600 & 2500 & 2016 & Yes \\
        \href{http://www.optil.io/}{Optil.io} & No & Eng & 3 & 11 & 300 & 2016 & Yes \\
        \href{https://www.topcoder.com/community/data-science/}{TopCoder Data Science} & No & Eng & 2 & 400 & - & 2001 & Yes \\
        \href{http://tunedit.org/}{TunedIT} & No & Eng,Pol & 3 & 36 & 10000 & 2008 & 2015 \\
        \hline
    \end{longtable}
}

Most likely, the best known is the \oj{Kaggle} platform \citep{Goldbloom_2010}, for which the primary objective is the organization of data-mining challenges with monetary prizes. However, Kaggle does not incorporate a regular online judge. Users of this service have to execute their code locally using test data provided by Kaggle and submit only the results generated by the algorithm. Providing the source code is not required. \oj{CrowdANALYTIX} implements an idea similar to Kaggle. The system hosts contests, and users submit outputs for a given problem before the deadline. The winner is chosen based on these solutions. \oj{DREAM Challenges} is another platform similar to Kaggle, but it focuses on problems related to systems biology and translational medicine \citep{Saez_Rodriguez_2016,Costello_2013}. As opposed to Kaggle, its primary objective is to solve scientific challenges; therefore, it is targeted specifically at researchers. It provides expertise and institutional support with the help of Sage Bionetworks, along with the infrastructure to host challenges via their Synapse platform \citep{Derry_2012}, creating a system with outstanding scientific value.

\oj{MLcomp} is another online platform that was designed with a slightly different idea in mind. It provides a cloud-based platform dedicated to data-mining research. Instead of challenges, it stores datasets. Any user can upload their own datasets and algorithms. All algorithms submitted by users are stored in the system and can be executed later on by any other user processing any uploaded dataset using the computing infrastructure provided by MLcomp. Such an approach creates a very versatile data analysis platform. Unfortunately, this platform is no longer maintained because its authors are currently developing a new online platform dedicated to planning and conducting research experiments called \oj{CodaLab}. However, there is another platform called \oj{OpenML} that offers a similar approach \citep{van_Rijn_2013}, providing at the same time a much more modern user interface design, as well as interfaces for the most popular scientific languages, namely R, Python, Java and supplemental REST API. Currently, it stores over 19,000 datasets.

It is also worth mentioning two other online platforms dedicated to publishing data-mining problems in the form of programming contests: \oj{TopCoder Data Science} (TopCoder DS) and \oj{TunedIT}. The former, previously called TopCoder Marathon Matches, has already organized several hundred competitions, including many sessions devoted to solving industrial-inspired challenges. Unfortunately, the set of supported programming languages is limited to only C++, C\#, and Java. The latter \citep{Wojnarski_2010}, as opposed to the TopCoder DS, obligates the users to submit reports describing their solutions. After the end of a particular challenge, all reports submitted by participants are verified by the organizers. Unfortunately, the website is not maintained as of late. Thus, new contests are not being added.

Finally, \oj{Optil.io} is a platform used to publish optimization problems that require the design of algorithms for optimizing objective functions on a provided data set \citep{Wasik_2016}. Users can submit solutions in several supported programming languages, as well as statically compiled Linux binaries. It is also possible to spread the code of the solution across several source files if a CMake file is also submitted by the user. Moreover, submissions can also use other libraries and specialized solvers that are supported by the proposed platform. The range of external software can be expanded on user demand. 

\subsection{Development platforms}

Anyone who would like to host a programming competition or a course using his/her own infrastructure can apply one of the several available online judge development platforms. These systems can be downloaded and installed locally, providing full administrative privileges to the user. Moreover, most of them can be adapted to user needs and integrated with external services. The list of services discussed in this subsection is presented in Table \ref{tbl:judges-development}.

{ \tiny
    \begin{longtable}{ cccccc }
    
        \caption{{Online judge systems that can be used as development platforms. For column descriptions, see Table \ref{tbl:judges-competitive}. Such systems usually do not provide problems database and do not require registration so the columns related to the number of problems and users have been removed.}}
        \label{tbl:judges-development} \\
        
        \hline \textbf{Name and URL address} & \textbf{OSS} & \textbf{GUI language} & \textbf{Compilers class id} & \textbf{Established} & \textbf{Active} \\
        \hline 
        \href{https://github.com/Aalto-LeTech/a-plus/}{A+} & Yes & Eng & 1 & 2017 & Yes \\
        \href{https://sourceforge.net/projects/cobalt/}{BOSS} & Yes & Eng & 3 & 2012 & 2009 \\
        \href{https://cloudcoder.org/}{CloudCoder} & Yes & Eng & 2 & 2012 & Yes \\
        \href{https://github.com/trampgeek/CodeRunner}{Code Runner for Moodle} & Yes & Eng & 3 & 2016 & Yes \\
        \href{http://www.domjudge.org}{DOMjudge} & Yes & Eng & 3 & 2004 & Yes \\
        \href{https://mooshak.dcc.fc.up.pt/}{Mooshak} & Yes & Eng & 2 & 2005 & 2015 \\
        \href{https://github.com/hit-moodle/moodle-local_onlinejudge}{Online Judge Plugin for Moodle} & Yes & Eng,Chi,Por,Pol & 3 & 2012 & 2015 \\
        \href{https://github.com/sio2project}{SIO2} & Yes & Eng,Pol & 2 & 2012 & Yes \\
        \href{http://testmycode.github.io}{TestMyCode} & Yes & Eng & 1 & 2013 & Yes \\
        \href{http://dsa.cs.tsinghua.edu.cn/oj/}{Tsinghua Online Judge} & No & Eng,Chi & 2 & 2012 & Yes \\
        \href{https://moodle.org/plugins/mod_vpl}{Virtual programming lab} & Yes & Eng & 3 & 2012 & 2015 \\
        \href{http://web-cat.org/home}{Web-CAT} & Yes & Eng & 3 & 2003 & Yes \\
        \href{http://dbis-group.uni-muenster.de/projects/xlx/}{xLx} & No & Eng & 1 & 2001 & 2008 \\
        \hline
    \end{longtable}
}
 
Currently, the platform that is the worthiest of recommendation is \oj{DOMjudge}. It is a fully automated judge system that allows users to prepare and perform programming contests following the rules defined for the ACM ICPC. It is an actively developed and feature-rich system. Its quality has been proven through its application in the ACM ICPC finals since 2012. \oj{Mooshak} \citep{Leal_2003} is another online platform designed for managing programming contests following rules similar to those designed for the IOI, as well as the ICPC. It offers a classic GUI, which has attained a very mature stage and is extensively used at many universities, for example, in Portugal. Finally, \oj{SIO2} is a stable online judge platform used and developed by the Polish Olympiad in Informatics.

There are also other development platforms that are dedicated to support of the educational process. For example, \oj{CloudCoder} \citep{Spacco_2015} is an open-source web-based system inspired by CodingBat that was designed to simplify the lives of instructors of programming courses by allowing them to assign exercises to students to assess their skills. \oj{Tsinghua University Online Judger} is a course-oriented online judge designed for universities. The system allows for the organization of programming classes using automatically evaluated challenges \citep{Zheng_2015}. It is hosted on the Tsinghua University server and allows users to add new problems. However, the source code of the system is not open source. \oj{BOSS} is a system that was designed to support the performance of programming courses using automatically judged assignments \citep{Joy_2005}. Over several years, it has evolved into a platform that supports the teaching of any topic, including the online verification of programming exercises. Unfortunately, the system was updated for the last time in 2009. Finally, \textbf{Web-CAT} is an  automated grading system implemented in Java to provide an approach to the evaluation of students' assignments. This is done by collecting test cases submitted by students for assessment using test coverage measures \citep{Edwards2008web}.

In the literature, one can also find online judge development platforms integrated with e-learning systems. One of the oldest such approaches is \oj{xLx}. It has existed since 2001, when the first prototype was developed to support the educational process conducted at WWU M\selectlanguage{ngerman}ü\selectlanguage{english}nster \citep{Husemann_2002}. It first published courses related to SQL and XML languages. In 2008, the second version of this platform was released; however, since that time, it has not been updated. Moreover, the source code of the system is not currently available. In 2011, Xavier and Coelho presented the review of several online judge platforms, as well as a description of the platform developed at the \oj{University of Porto} inspired by the Moodle and DOMjudge systems \citep{Xavier_2011}. Kaya and \selectlanguage{ngerman}Ö\selectlanguage{english}zel in \citep{Kaya_2014,ozel2012online} presented a similar approach, but enriched with a plagiarism detection package called \oj{Moss} \citep{Schleimer_2003}. \oj{Code Runner} is a question type plug-in for Moodle that allows one to execute code submitted by a particular student as an answer for a broad range of programming puzzles, and it supports various programming languages simultaneously. It is intended primarily for use in computer programming courses, although it can be used to grade any question for which the answer is represented in textual form. Currently, the most comprehensive and mature plug-in for Moodle is \oj{Virtual Programming Lab} \citep{rodriguez2012virtual} which allows users to edit their source code interactively in a browser, develop custom grading scripts, and detect plagiarism automatically. Unfortunately, it has not been updated for over a year. Thus, it is not compatible with the newest version of Moodle. Finally, the \oj{Online Judge plug-in for Moodle} \citep{Zhigang_2012} is a component that allows users to submit solutions to programming exercises and supports approximately 40 programming languages. However, it uses only the deprecated activities API supported by Moodle up to version 2.2 and can currently only be used in legacy mode. Because Moodle is a very popular online system supporting the educational process, there have been additional attempts to integrate it with online judge systems. However, all of them are either currently inactive, not compatible with the current stable version of Moodle or exist as proof-of-concept projects only, e.g., \citep{Danutama_2013,MorenoCarral2013,Davuluri2016}. A similar plug-in-based concept is provided by \oj{TestMyCode}, which allows for an automated transfer of programming assignments directly from within the most widely known IDEs \citep{Parte2013}. It is worth noting that the submission phase is much simpler in this case; nevertheless, the evaluation is provided based on predefined unit tests. Moreover, a new e-learning platform called \oj{A+} has been recently published by Aalto University based on previous work by \citep{Karavirta2013}. The main advantage lies in its flexible utilization of external services. Additionally, it provides a REST API, allowing for the automatic acquisition of various information stored in the system.

\section{Evaluation methodology}
\label{sec:evaluation-methodology}

The most important component of each online judge system is an evaluation engine that assesses submissions uploaded by users. The evaluation procedure of such systems usually consists of three steps described in the definition \ref{def:evaluation-procedure}. The first step is relatively simple and technical. First, it usually relies on the execution of an appropriate compiler. Second, it verifies that the compiler does not return any compilation errors, and compilation time and output binary size do not exceed the specified threshold. Using recursive macro-definitions, it is relatively easy to abuse the system this way. 

Usually, when the solution compiles successfully, there are no obstacles preventing it from execution. However, sometimes online judge systems allow for submitting static binaries compiled by the author on the client side. In such a case, the system must verify that all required, dynamically linked libraries are available, and that the binary is compatible with the evaluation environment's architecture.

The remaining steps of the evaluation procedure are more complex and diversified. We review them in detail in Sections \ref{ssec:assessment-phase} and \ref{ssec:scoring}. However, before presenting our explanation, we shall introduce basic concepts related to the theory of combinatorial problems (see section \ref{ssec:optimization-problems}). As most problems published on online judge systems are some type of combinatorial problem, such explanation is crucial to better understand the evaluation process.

\subsection{Combinatorial problems}
\label{ssec:optimization-problems}

Any combinatorial problem $\Pi$ is defined in the form of a set of parameters associated with it and a set of required constraints that must be satisfied by its solution. An instance $I$ of the problem $\Pi$ is obtained by setting particular values for all parameters considered in the problem definition. One of the most well-known combinatorial problems is the 0-1 knapsack problem, which is defined by a given set of $n$ items, each described by a weight $w_{i}$ and a value $v_{i}$, along with a maximum weight capacity $W$. The solution of this problem is a combination of items (assuming that every item can be introduced into the knapsack at most once) that fit in the knapsack for which the total value is maximal. The set of all instances of the problem $\Pi$ is called the domain of the problem $D_{\Pi}$.

In fact, two main classes of combinatorial problems can be distinguished, i.e., decision problems and search problems. 

\begin{definition}[Decision problem]
A decision problem $\Pi$ is a problem for which a solution is an answer either yes or no for a question associated with the set of its parameters. 
\end{definition}

An example of this problem can be a particular instance of the knapsack problem and the following question associated with it: is it possible to pack the knapsack according to a given set of items reaching a total value not lower than a particular constant $V_{max}$ and, at the same time, not exceeding the maximum weight capacity of knapsack $W$.

\begin{definition}[Search problem]
A search problem is a problem for which a solution is either a particular object (or value) satisfying given constraints or the answer no when such a solution does not exist.
\end{definition}

In this case, the problem relies on finding any feasible solution of the knapsack problem based on a given set of items, i.e., such a combination of items that fit in the knapsack for which total value will not be lower than the particular constant $V_{max}$.

An \textit{optimization problem} is a special case of the search problem where the solution is represented by the optimal object (or value) according to a given objective function. An example is another version of the knapsack problem formulated in the following manner: find such a combination of items that fit in the knapsack for which the total value is maximal. In contrast to the aforementioned search problem, here the solution is the best one among a set of all feasible solutions for this problem. 

For more information regarding optimization problems and their complexity, see the excellent book by Garey and Johnson \citep{Garey1979}.

\subsection{Assessment}
\label{ssec:assessment-phase}

The assessment phase is usually the most computationally expensive. It requires execution of the compiled solution in each considered test instance. Each of these executions can take even several dozen minutes for complex industry-inspired problems. Below, we provide the formal definition of a test instance:

\begin{definition}[Test instance]
Let $\Sigma$ denote an alphabet used to encode both input and output data. Test instance $t_i \in T$, where $T$ is a set of all considered for the particular problem test instances, is defined as a triple $t_i = (d_i,o_i,p_i)$, where $d_i = \Sigma^*$ is an input data, $o_i = \Sigma^*$ is a reference output data, and $p_i$ is a set of parameters passed to the evaluation engine.
\end{definition}

In the vast majority of problems, the alphabet $\Sigma$, that is used to encode both input and output data, consists of digits, spaces, and new line characters $\Sigma=\{0, 1, \dotsc, 9, \texttt{\textvisiblespace}, \backslash n\}$. Sometimes, it is extended with lower and upper case letters and up to several special characters. Nevertheless, the format used to encode both types of data is usually as simple as possible for parsing purposes. Usually, it is represented by a list of integer numbers of which structure is defined in the problem description or a simple comma-separated values (CSV) file often used in data-mining applications. Commonly, each input and output data are represented by a single file redirected to the standard input and from the standard output of the solution program, respectively. However, in case of specific problems, it could happen that several files stored in the single archive are provided.

When, according to the problem definition, for each input instance $d_i$ there is exactly one correct solution, this solution can be precomputed and stored in the reference output data file $o_i$. This way, the evaluation engine computes the reference solution immediately, and next uses it during verification of each user submission. Whenever there are many feasible solutions for a single input instance, the reference output data file can store certain precomputed values that allow the computational complexity of the evaluation process to be reduced. When no additional data can simplify the evaluation process, the reference output data $o_i$ can be empty (i.e., $o_i = \varnothing$).

The set of parameters, $p_i$, represents the specific resource limitations (e.g., CPU time, maximum utilization of RAM) that cannot be exceeded during the evaluation based on this particular instance. However, additional parameters can also be passed if necessary, such as random number generator seed when the engine allows randomized solutions, or a maximum size limit for output data generated as a result of the solution execution for this particular instance. When the evaluation engine is configured to use the default resource limits and no other parameters are needed, the set $p_i$ can be empty (i.e., $p_i = \varnothing$).

During the assessment phase, the solution compiled as the output of the submission phase is used. Below, we provide the formal definition of the solution:

\begin{definition}[Solution]
A solution is a function, $b(d_i,p_i^\prime)\to o_i^\prime$, representing the binary form of the submission that, based on the input data $d_i$, computes output data $o_i^\prime$, taking into consideration execution parameters $p_i^\prime$ provided by the evaluation engine.
\end{definition}

The set of execution parameters provided by the evaluation engine $p_i^\prime$ can be equal to the set of parameters defined as part of the particular test instance (i.e., $p_i^\prime = p_i$). However, in particular, $p_i^\prime$ can be the subset of $p_i$ (i.e., $p_i^\prime \subset p_i$), $p_i^\prime$ can be a completely different set of parameters or $p_i^\prime$ can be empty (i.e., $p_i^\prime = \varnothing$). As opposed to $p_i$, the set $p_i^\prime$ is usually empty, because parameters influencing the evaluation process are hidden from the solution. The most commonly used execution parameters often passed to the user's solution are the seed for the random number generator and maximal time limit that cannot be exceeded by the solution during its execution for the particular test instance.

\begin{definition}[Evaluation engine]
An evaluation engine is a function, $E(b,t_i)\to (s_i,v_i,e_i)$, that executes the binary file $b$, giving it, as an input, the test instance $t_i$, and returns the execution status $s_i$, an evaluation score calculated for the solution output, $v_i\in \mathbb{R}$, and the list of statistics collected for the execution process $e_i$.
\end{definition}

The status $s_i$ of the submission execution can take one of the following values:
\begin{itemize}
  \item Accepted (ACC) when the submission execution terminates successfully without any runtime error, without exceeding resources limits, and returns feasible output data coherent with the format described in the problem description;
  \item Time Limit Exceeded (TLE), means incorrect execution of the submission due to exceeding the maximal processing time limit; 
  \item Memory Limit Exceeded (MLE), means incorrect execution of the submission due to exceeding the maximal RAM utilization limit (covering both the stack and heap); 
  \item Wrong Answer (WA), means that the program generated an output of unknown format (i.e., the existing format is not coherent with the format requested in the problem description) or some additional constraints formulated in the problem description were not satisfied; 
  \item Runtime Error (RE), means that a runtime error occurred during the particular submission's execution;
  \item Output Limit Exceeded (OLE), means that the submission exceeded the maximal limit for the size of output data.
\end{itemize}
In the case of a WA status code, the user can also receive additional information regarding the reason the answer is incorrect. However, it is not a common practice. Various online judges can introduce other statuses. However, the aforementioned are the most commonly used. Most commonly, users can instantly see the status of the submission, at least, for some example instances, as soon as it is assessed. 

Many online judge systems follow the ICPC rules and evaluate the submissions on each test instance in a binary way---as a correct or incorrect solution only. In such a case, the evaluation score is always equal to 0 (i.e., $v_i=0$). There are two major cases when this value is used. First, in optimization problems, when it stores the value of an objective function computed for the output data, obtained as a result of a particular submission execution on a particular test instance.  It can be both, a classical optimization problem \citep{Garey1979} or a code-golf problem, when the objective is to optimize the size of the source code solving some task. Second, in competitions following IOI rules, $v_i$ characterizes the score that the user gains for receiving the Accepted status for this particular instance (i.e., $s_i=ACC$). In general, different test instances can be characterized by various score values or even more complex scoring procedures can be also applied. For example, Polish Olympiad in Informatics penalizes solutions that use more than the half of the time limit by decreasing the score proportionally, according to the following formula:
\begin{equation}
v_i = V_i \cdot \min\left(1.0, 2.0\cdot\frac{\mathcal{T}_i-\tau_i}{\mathcal{T}_i}\right)
\end{equation}
Here, $V_i$ denotes the maximal number of points that can be awarded for some test instance, $\mathcal{T}_i$ denotes the maximal time limit for the solution, and $\tau_i$ denotes the processing time used by the solution to generate the output for the particular instance.

Finally, statistics $e_i$, collected for the execution process of the solution, usually include information about maximum values of resource utilization observed during the particular submission execution (e.g., time and memory consumption). In particular, when the online judge does not share such information with the user, the list of execution statistics can be empty (i.e., $e_i=\varnothing$). 

\subsection{Scoring}
\label{ssec:scoring}

The objective of the scoring phase is to compute an aggregated status $s$ and aggregated evaluation score $v$ of each user submission. These values are then used to display results of evaluation procedure to the user and rank solutions of the problem.

If and only if, for all test instances, the solution received ACC status, then the aggregated status is also ACC. Otherwise, the most common procedure is to select the first status different than ACC:
\begin{align}
s &= ACC \Leftrightarrow \forall_i s_i = ACC \\
s &= s_j \Leftrightarrow (\forall_{i<j} s_i = ACC) \land (s_j \neq ACC)
\end{align}
However, some variations are sometimes utilized, for example, returning an RE status when any Runtime Error occurs before any other statuses.

Computing the aggregated evaluation score $v$ is relatively simple when the system does not deal with optimization problems. In such a case, for systems evaluating submissions in a binary way, following ICPC rules, this score is always equal to 0 (i.e., $v=0$). Otherwise, it is usually the total evaluation score computed for all correctly solved instances:
\begin{equation}
v = \sum_{i=1}^{|T|} \begin{cases}
v_i, & \text{if } s_i = ACC \\
0, & \text{otherwise}
\end{cases}
\end{equation}

In the case of optimization problems, the system has to consider the score computed based on the best solution among all submissions or a reference solution computed by the problem author. In most cases, the following formula is used to rank the submitted solutions for problems where the objective function is maximized:
\begin{equation}
\label{eqn:score-max}
v = \frac{100}{|T|}\sum_{i=1}^{|T|}\begin{cases}
\frac{v_i}{b_i}, & \text{if } s_i = ACC \\
0, & \text{otherwise}
\end{cases}
\end{equation}
where $b_i$ denotes the best solution score for the $i$-th instance (i.e., the solution among all submissions for which the objective function score computed for this particular instance has the optimal value). Similar types of formulas can be easily proposed for minimization problems as well, taking into account an extended scoring procedure and execution times or, even, the utilization of other computing environment resources.

\subsection{Online judge}

Based on the definitions presented in this section we can define an online judge system as follows:

\begin{definition}[Online judge system]
\label{def:online-judge}
An online judge system is an online service that performs any of the steps of the evaluation procedure in a cloud, i.e.:
\begin{enumerate}
\item collects, compiles sources if needed, and verifies executability of resultant binaries,
\item assesses solutions $b(d_i,p_i^\prime)$ based on a set of test instances, $T$, defined for a particular combinatorial problem $\Pi$ in a reliable, homogeneous, evaluation environment,
\item computes the aggregated status $s$ and evaluation score $v$ based on the statuses and scores for particular instances (i.e., $s_i$ and $v_i$, where $1 \leq i \leq |T|$).
\end{enumerate}
\end{definition}

\section{Example application based on Optil.io platform}
\label{sec:results}

\begin{figure}[ht!]
\begin{center}
\includegraphics[width=0.7\columnwidth]{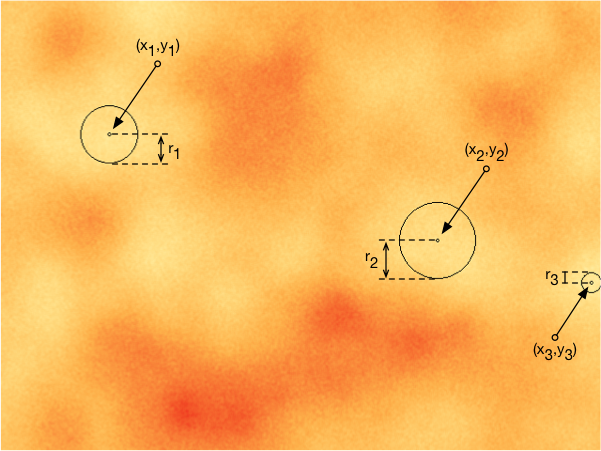}
\caption{{\label{fig:instance}
Visualization prepared for an example input instance of the first challenge conducted on the Optil.io platform. The objective was to find locations $(x_i,y_i)$ for three factories with various radii $r_i$. Example positions of factories and locations that they affect are marked with black circles. The close-to-red color displayed on the map represents the areas where the values of the discontent coefficient are higher. The better locations for factories are represented by circles surrounding lighter areas.%
}}
\end{center}
\end{figure}

To demonstrate the process of an online judge system, in this section, we would like to present the results of an example contest that we conducted using the Optil.io online judge system \citep{Wasik_2016} described in section \ref{ssec:data-mining}. We selected this platform because, as its authors, we had the biggest control over the system. Based on this platform, we organized a contest where the example, an optimization-based challenge, was solved. It was a variation of the multiple facilities location problem, for which the computational complexity was proven to be NP-hard \citep{Megiddo_1982}. Contestants were responsible for placing several factories on the Euclidian plane, ensuring minimization of the total discontent of people living next to them. Each factory was described by a single parameter, i.e., influence range. In turn, each position on the plane was characterized by the dissatisfaction coefficient computed for this position when it was affected by any factory. The visualization prepared for an example input instance of the conducted challenge and its solution is presented in Figure \ref{fig:instance}. Participants had to solve the problem for ten randomly generated instances by placing from 3 to 50 factories on a grid of up to 1000-by-1000 points. For each considered instance, the processing time limit could not exceed 10 seconds, and memory used could not exceed 1 GB. For each submission executed on a particular test instance, the system generates one of the execution status codes described in Section \ref{ssec:assessment-phase} and computes a score using Equation \ref{eqn:score-max}.

We recruited the participants of this challenge primarily among a large group of second-year students of a Computer Science course at the Poznan University of Technology. Before participation in the competition, all of them accepted the terms of use of the tested service and the cookies policy according to European law. Before conducting statistical analysis of this study's results, we anonymized all data in accordance with the previously accepted terms of use.

The contest held lasted 18 days, from January 23rd to February 9th, 2016. However, after that time, the problem was still available on the website. During the period of the contest, 85 users submitted 1191 solutions to the problem, and since its end, another 400 answers have been submitted. It is worthy to underline that the main aim of the contest was to identify the best solution. It is typical for industry-inspired challenges that the solution is extremely important; however, the winning person is not that important. Therefore, the participants could refine their solutions during the lasting competition motivated by the online rank list presenting the best solution of each contestant evaluated on each instance. Such an approach allows utilization of crowdsourcing to generate much better solutions than during challenges when participants submit a single solution at the end of the contest. The objective function ratio of the best solution in the current time to the best counterpart provided in the entire competition and the number of users participating in the contest over the days is presented in Figure 2. At the beginning of the competition, the solutions submitted by users were over six times worse than the one that won it. However, in a relatively short amount of time, (i.e., four days), the best solution was significantly improved, i.e., the best score in that time was two times better than the previous one. This is an expected behavior, because at the beginning of the optimization, the best of submitted solutions is usually relatively far away from the global optimum, such that it can be easily improved by the contestants. In the middle of this contest, (i.e., a week since the beginning of the contest), the best solution score was close to the contest winner. It is worth mentioning that the breakthrough in the solution space was obtained by a small number of users, and then the final tuning of this solution occurred when the number of active users remarkably increased. The winning solution was submitted 52 minutes before the end of this contest. Despite that, the winning solution was significantly better than the initially submitted solution from January 23rd, and the improvement during the second half of the contest, between February 2nd and 13th, was not so significant. Nevertheless, it was still equal to almost 13\%. It is worth underlining that, while optimizing industrial processes, e.g., in logistics, such a decrease of costs can potentially generate savings on the order of millions of dollars.

In turn, Figure \ref{fig:correct} presents the number of correct and incorrect solutions submitted by users. The solution was classified as incorrect when it generated the wrong answer or raised execution errors (i.e., TLE, MLE, RE, or WA; see section \ref{ssec:assessment-phase}). At the beginning of the competition, the number of submitted solutions classified as potentially incorrect significantly overwhelmed the number of correct ones. However, the contestants quickly learned how to successfully submit their solutions, and thus at the end of the competition, the number of submitted solutions classified as potentially correct oscillated approximately 50\%. Such an observation related to the ratio of correct and incorrect solutions confirmed the results characterizing other online judge systems reported by Manzoor \citep{manzoor2006analyzing}. This proves that the proposed GUI is ergonomic as well as user-friendly. The ca. 60\% of invalid submissions presented in Figure 3 is primarily the consequence of two reasons. First, usually, users do not test their code sufficiently before submission. They assume that this is a role of the online judge system to verify their code and submit an algorithm that generates incorrect answers for some instances or even does not compile. Second, the users do not comply with the execution limitations described in the challenge definition. Such solutions are terminated by the online judge system, receiving TLE status at the same time.

After five days of the competition, the observed improvement slowed down significantly unless the significantly larger number of participants joined the competition. This observation demonstrates perfectly the advantage of crowdsourcing. A crowd of highly skilled participants can achieve much better results quality than a single developer or even a small team. The results of the contest organized using Optil.io platform proved that it could be successfully applied to solve optimization problems interactively by a large number of programming enthusiasts.

\begin{figure}[ht!]
\begin{center}
\includegraphics[width=0.7\columnwidth]{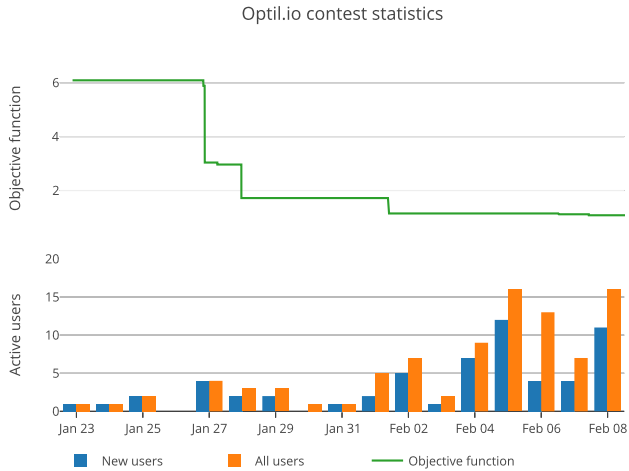}
\caption{{\label{fig:solution-val}
Upper plot: Objective function ratio computed between the best solution currently and the winner of the entire competition. The objective function score is averaged over all successfully processed test instances. Lower plot: The number of users who submitted their solutions during that particular day. The orange bars denote the total numbers of users; the green bars denote only new users who submitted their solutions on that particular day for the first time.
}}
\end{center}
\end{figure}

\begin{figure}[ht!]
\begin{center}
\includegraphics[width=0.7\columnwidth]{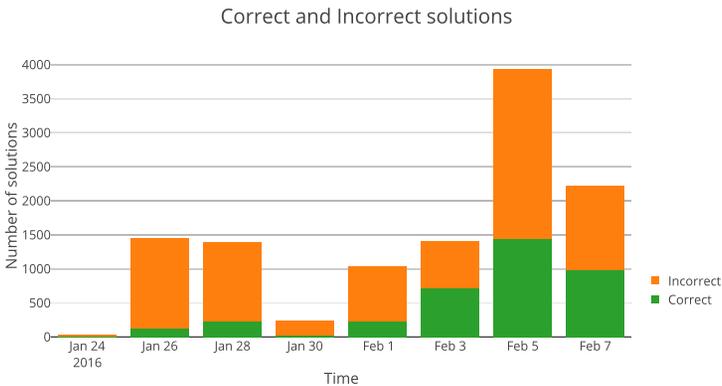}
\caption{{\label{fig:correct}
The numbers of correct and incorrect solutions submitted during the consecutive days of the contest.%
}}
\end{center}
\end{figure}

\section{Conclusion}

Online judge systems have a very long history, lasting over half of a century. However, their popularization and widespread use have been empowered over the first decade of the twenty first century. These systems have emerged as simple web applications provided by universities to support lecturers in teaching, and students in preparation and participation in competitive programming championships. Meanwhile, the organizations hosting such programming competitions became interested in the application of online judge systems to share challenges formulated in the past to help users in preparation for upcoming events. It caused that many online platforms emerged publishing challenges originated from competitive programming contests. The primary reason was the rapid spread of Internet access, which gave many contestants worldwide an opportunity to easily submit and verify their solutions. 

A majority of these systems are open for users worldwide, who want to improve their algorithmic and programming skills. Moreover, there are few that support the organization of customized contests addressed to the public or a limited group of participants, solving challenges originating from local, as well as remote archives. To strengthen the popularity, several systems provide independent, localized instances for particular nationalities. The contest organization procedure often varies among the systems in a sense of time duration and number of challenges considered in a single round, number of provided programming languages, and even participant groups. However, in many of them, the inspiration with ICPC or IOI contest is noticeable. The contests are usually organized annually, a few times a year, or on demand of the challenge organizer. For example, Facebook and Google, organize their own competitions annually. Sometimes, the participants are scored, not only for the quality of their solutions, but also for successful identification of issues in other participant submissions. Usually, to distinguish submissions that are characterized by the same quality, the final score considers also the submission time or even the number of participants or groups that have submitted that solution previously. There are also other websites that present less classical applications of online judges designed to host domain-specific challenges or support only a textual representation of the output for computationally intensive problems. It is worthy to note that there exist systems that support the selection of appropriate problems within a tremendous set of published challenges, or even meta systems that are integrated with many online judges allowing the user to submit his/her solution from a single place. A distinguishable feature of competitive programming systems is usually a simple GUI that only supports the most popular, low-level programming languages such as C/C++ and Pascal.

Online judges are also used in rather more sophisticated applications, such as education, employee recruitment and even data mining. Online judge systems are of interest for educational organizations because they support fully-automatic and accurate evaluation of student assignments. Therefore, the teacher can focus on exercise quality and the teaching process. We previously mentioned certain successful applications of these systems in the teaching of IT courses at various universities. Currently, the systems that allow users to strengthen their commitment and excitation, by incorporation of additional elements of gamification and social networking, are significantly gaining popularity. For example, students are strengthening their coding skills, not by solving typical exercises, but by developing specialized codes to protect their own units or attack enemies in a multiplayer strategy game. On the other side, there are also tools providing programming courses and professional tutorials for computer science enthusiasts, for example, cracking passwords using the assembly language. It is worth remarking that educational online judge systems usually provide a rich, ergonomic GUI and gamification elements to engage students in the problem-solving process. The number of programming languages supported by such systems is also much higher, in comparison to competitive programming systems, so as to focus mainly on the educational process, independently of the programming language that is taught.

To ensure completeness of presented classifications, we also considered online compilers because they share common mechanisms used to compile and execute source code submitted by the user with online judge systems. Moreover, they are often used as a crucial component in the development of online judge systems. In this group, we can find a wide range of tools, from simple online services supporting compilation and execution of user-submitted codes, to systems providing cloud-based IDEs that support collaboration in real-time by sharing a dedicated virtual development environment; however, a common feature is the ability to support users in solving programming assignments using only a web browser, without access to the standalone software. Such systems are especially useful for users who usually work with various workstations without administrative privileges.

The platforms proposed to support recruitment process usually integrate and extend basic mechanisms provided by online judge systems. These systems allow developers to strengthen their skills by competing in various programming competitions, thus, building their professional reputation within the community. They are also very useful for recruiters who can easily find many reliable employment candidates. A few of them allow even for delivery of fully-customized challenges for specific candidates and their assessment in real-time. All these systems are characterized by a professional GUI and enhanced functionality that allow for an analysis of programming skills based on traced interactions of the person being recruited using the system. Supported programming languages are usually limited to the most popular in the industry, such as C/C++, C\#, Java, JavaScript, Python, PHP.

There are also successful applications of online judge systems in platforms supporting the organization of data mining challenges. They are dedicated to solving complex industry- as well as science-inspired problems. The common usage scenario for these systems is that users submit the output of their programs obtained for specific test instances that are then assessed by the particular challenge organizers, and a final score is computed. However, it is worthy to note that the user solutions can be rarely submitted in other forms, for example, by providing the algorithm's source code, or even specialized reports summarizing the result. The number of challenges provided by these systems is growing extremely fast.

The approach offered by online judge platforms has such a large impact that the concept they utilize has already been named cloud-based evaluation, or, strictly following the cloud computing naming scheme, Evaluation-as-a-Service (EaaS). At least two international workshops devoted to this topic were organized in 2015: Workshop on cloud-based evaluation approaches in the United States \citep{Muller_2016} and Evaluation-as-a-Service Expert Workshop in Switzerland \citep{Hopfgartner_2015}. There is also the Metaheuristics in the Large initiative that integrates around this topic many prominent researchers from the field of operational research \citep{Swan_2015}. The objective is to provide computing infrastructure that allows enthusiasts or researchers to incorporate state-of-the-art metaheuristics and solvers into an optimization workflow. 

Unfortunately, most of the aforementioned systems are closed source. Thus, it is hard to apply them successfully, considering various user expectations. That is why, at last, but not least, we mentioned the systems that can be downloaded and installed on the user's own infrastructure by practitioners to provide an instance of a highly-configurable online judge system. This can be easily applied to the organization of their own programming competition, as well as support the educational process by the automated evaluation of student programs. Thus, such systems are of great interest for universities and organizations providing e-learning courses.

Currently, the systems providing problem archives for participants of competitive programming contests constitute the majority of all online judge systems because the development of these systems started approximately 10 years earlier than the remaining ones. However, in the near future, this proportion will certainly invert as a result of the intensive development of cloud-based computing \citep{Puthal_2015}, which forms the basis for systems designed with EaaS architecture in mind.

The classification proposed in this article is addressed to all interested practitioners to support in finding the appropriate online judge platform regardless of their needs. A classification is the most important contribution distinguishing this survey from earlier reviews. For example, Ihantola et al. described principles that are crucial in the design of online judge systems just for assessment of programming assignment \citep{Ihantola_2010}. Another example is the article by N{\'e}meth and L{\'a}szl{\'o}, which focused on the classification of online judge systems that only supports the organization of programming contests \citep{Nemeth_2015}.

Moreover, we formally defined what an online judge system is, and described an evaluation procedure commonly used by such systems. While the intuitive perception of an online judge system was first introduced by Kurnia et. al in 2001 \citep{Kurnia_2001}, a formal definition of such a system remained undefined until hitherto. Similarly, the evaluation procedure mentioned several times, especially in relation to ACM ICPC and IOI contests \citep{cormack2006structure}, was only described informally, with an emphasis on the context of the particular programming contest.

Additionally, we presented an example use case based on the Optil.io platform, which is the first system, worldwide, providing a continuous evaluation of algorithms solving complex optimization challenges in a safe and homogenous cloud-based infrastructure. Optimization problems are problems significant from the point of view of possible applications as well as often difficult to solve problems that are formulated in a very broad range of fields from logistics \citep{Silva_2008} and decision making \citep{Marler_2004,Blazewicz_2011,Blazewicz_2014}, through software design \citep{Marszalkowski_2015}, to even biology \citep{Wang_2010,Szostak2016a,Prejzendanc2016,Antczak_2016,Lukasiak_2015,Wasik2013a}. Despite the huge development of mathematical programming solvers and meta-heuristics that have occurred in recent years, many of these problems are still difficult to solve. These are NP-hard problems that can be solved by currently known methods for small instances or some special cases only. According to the ``no free lunch'' theorem applied for optimization purposes \citep{Wolpert_1997}, designing well-performing algorithms for such problems requires the application of sophisticated approaches for each of them. The experiment conducted using the Optil.io platform demonstrated that online judge systems can significantly enhance this process by providing a reliable platform that can be applied to effectively discovery the solutions of science- and industry-inspired optimization problems using a crowdsourcing approach. We are sure that Optil.io and other online judge systems can assist users in utilizing the power of crowdsourcing to solve the most difficult problems known in science and industry.

\bibliographystyle{ACM-Reference-Format}
\bibliography{survey.bib%
}

\newpage
\section*{Electronic supplement}

\subsection*{URL addresses of considered systems}

{ \tiny
    \begin{longtable}{ cccccc }
    
        \caption{List of online judge systems presented in the article in Tables \ref{tbl:judges-competitive} to \ref{tbl:judges-development} together with URL addresses published to allow printing of URLs.}
        \label{tbl:judges-url} \\
        
        \hline \textbf{Name} & \textbf{URL address} \\
        \hline 
        A2 Online Judge & \href{https://a2oj.com/}{https://a2oj.com/} \\
        AC 2333 & \href{https://ac.2333.moe/}{https://ac.2333.moe/} \\
        AcDream & \href{http://acdream.info/}{http://acdream.info/} \\
        ACM-ICPC live archive & \href{https://icpcarchive.ecs.baylor.edu}{https://icpcarchive.ecs.baylor.edu} \\
        ACM-Kyrgyzstan Subregion & \href{http://www.olymp.krsu.edu.kg/GeneralProblemset.aspx}{http://www.olymp.krsu.edu.kg/GeneralProblemset.aspx} \\
        Adjule Online Judge & \href{https://adjule.pl/}{https://adjule.pl/} \\
        Aizu Online Judge & \href{http://judge.u-aizu.ac.jp/onlinejudge/}{http://judge.u-aizu.ac.jp/onlinejudge/} \\
        Al Zimmermann's Programming Contests & \href{http://www.azspcs.net/}{http://www.azspcs.net/} \\
        BNU OJ & \href{http://www.bnuoj.com/v3/index.php}{http://www.bnuoj.com/v3/index.php} \\
        Carribean Online Judge & \href{http://coj.uci.cu/index.xhtml}{http://coj.uci.cu/index.xhtml} \\
        CDOJ & \href{http://acm.uestc.edu.cn}{http://acm.uestc.edu.cn} \\
        Codeforces & \href{http://codeforces.com/}{http://codeforces.com/} \\
        Don Mills Online Judge & \href{https://dmoj.ca/problems/}{https://dmoj.ca/problems/} \\
        e-olymp & \href{http://e-olymp.com}{http://e-olymp.com} \\
        EI Judge & \href{http://acm.mipt.ru/judge}{http://acm.mipt.ru/judge} \\
        Facebook Hacker Cup & \href{https://www.facebook.com/hackercup}{https://www.facebook.com/hackercup} \\
        Fuzhou University Online Judge & \href{http://acm.fzu.edu.cn/}{http://acm.fzu.edu.cn/} \\
        Google Code Jam & \href{https://code.google.com/codejam/}{https://code.google.com/codejam/} \\
        Herbert Online Judge & \href{http://herbert.tealang.info/problems.php}{http://herbert.tealang.info/problems.php} \\
        HIT ACM/ICPC & \href{http://acm.hit.edu.cn/hoj}{http://acm.hit.edu.cn/hoj} \\
        HUSTOJ & \href{http://acm.hust.edu.cn/}{http://acm.hust.edu.cn/} \\
        Indian Computing Olympiad Problems Archive & \href{http://opc.iarcs.org.in/index.php/problems/}{http://opc.iarcs.org.in/index.php/problems/} \\
        Internet Problem Solving Contest & \href{http://ipsc.ksp.sk}{http://ipsc.ksp.sk} \\
        Light OJ & \href{http://lightoj.com}{http://lightoj.com} \\
        LYDSY & \href{http://www.lydsy.com/JudgeOnline/}{http://www.lydsy.com/JudgeOnline/} \\
        Main & \href{http://main.edu.pl}{http://main.edu.pl} \\
        National Taiwan University Online Judge & \href{http://acm.csie.ntu.edu.tw/ntujudge/index.php}{http://acm.csie.ntu.edu.tw/ntujudge/index.php} \\
        National Tsing Hua University Online Judge & \href{http://acm.cs.nthu.edu.tw/}{http://acm.cs.nthu.edu.tw/} \\
        North University of China Online Judge & \href{https://noj.io/}{https://noj.io/} \\
        P3G & \href{http://wcipeg.com/}{http://wcipeg.com/} \\
        Peking University Judge Online & \href{http://poj.org/}{http://poj.org/} \\
        Petrozavodsk State University & \href{http://acm.petrsu.ru}{http://acm.petrsu.ru} \\
        Project Euler & \href{https://projecteuler.net/}{https://projecteuler.net/} \\
        SPOJ & \href{http://www.spoj.com/}{http://www.spoj.com/} \\
        SPOJ PL & \href{http://pl.spoj.com/}{http://pl.spoj.com/} \\
        Szkopuł & \href{http://szkopul.edu.pl/}{http://szkopul.edu.pl/} \\
        Teddy Online Judge & \href{https://www.teddyonlinejudge.net/}{https://www.teddyonlinejudge.net/} \\
        Timus Online Judge & \href{http://acm.timus.ru/}{http://acm.timus.ru/} \\
        TJU ACM-ICPC Online Judge & \href{http://acm.tju.edu.cn/toj/}{http://acm.tju.edu.cn/toj/} \\
        TopCoder Competitive Programming & \href{https://www.topcoder.com/community/competitive\%20programming/}{https://www.topcoder.com/community/competitive\%20programming/} \\
        USA Computing Olympiad & \href{http://usaco.org}{http://usaco.org} \\
        UVa Online Judge & \href{http://uva.onlinejudge.org/}{http://uva.onlinejudge.org/} \\
        CrowdANALYTIX & \href{https://www.crowdanalytix.com/}{https://www.crowdanalytix.com/} \\
        DREAM Challenges & \href{http://dreamchallenges.org/}{http://dreamchallenges.org/} \\
        Kaggle & \href{http://www.kaggle.com}{http://www.kaggle.com} \\
        MLcomp & \href{http://mlcomp.org/}{http://mlcomp.org/} \\
        OpenML & \href{http://www.openml.org/}{http://www.openml.org/} \\
        Optil.io & \href{http://www.optil.io/}{http://www.optil.io/} \\
        TopCoder Data Science & \href{https://www.topcoder.com/community/data-science/}{https://www.topcoder.com/community/data-science/} \\
        TunedIT & \href{http://tunedit.org/}{http://tunedit.org/} \\
        CheckiO & \href{http://www.checkio.org}{http://www.checkio.org} \\
        Code Fights & \href{https://codefights.com/}{https://codefights.com/} \\
        Codeboard & \href{https://codeboard.io/}{https://codeboard.io/} \\
        Codecademy & \href{https://www.codecademy.com/}{https://www.codecademy.com/} \\
        CodeChef & \href{http://www.codechef.com/}{http://www.codechef.com/} \\
        CodeHunt & \href{http://www.codehunt.com}{http://www.codehunt.com} \\
        Codewars & \href{http://www.codewars.com}{http://www.codewars.com} \\
        CodinGame & \href{http://www.codingame.com/start}{http://www.codingame.com/start} \\
        CodingBat & \href{http://codingbat.com/}{http://codingbat.com/} \\
        Embedded Security CTF & \href{https://microcorruption.com}{https://microcorruption.com} \\
        Exercism & \href{http://exercism.io}{http://exercism.io} \\
        Jutge.org & \href{https://jutge.org/}{https://jutge.org/} \\
        Leek Wars & \href{https://leekwars.com/}{https://leekwars.com/} \\
        LeetCode Online Judge & \href{http://leetcode.com/}{http://leetcode.com/} \\
        Programming Grid & \href{http://programming.grids.cn/programming/}{http://programming.grids.cn/programming/} \\
        Python Challenge & \href{http://www.pythonchallenge.com/}{http://www.pythonchallenge.com/} \\
        RACSO & \href{https://racso.lsi.upc.edu/juez/}{https://racso.lsi.upc.edu/juez/} \\
        The AI Games & \href{http://theaigames.com/}{http://theaigames.com/} \\
        URI Online Judge & \href{https://www.urionlinejudge.com.br/judge/login}{https://www.urionlinejudge.com.br/judge/login} \\
        C++ Shell & \href{http://cpp.sh/}{http://cpp.sh/} \\
        Codeanywhere & \href{https://codeanywhere.com}{https://codeanywhere.com} \\
        Codepad & \href{http://codepad.org/}{http://codepad.org/} \\
        CodeSkulptor & \href{http://www.codeskulptor.org/}{http://www.codeskulptor.org/} \\
        Coding Ground & \href{http://www.tutorialspoint.com/codingground.htm}{http://www.tutorialspoint.com/codingground.htm} \\
        Codio & \href{http://codio.com}{http://codio.com} \\
        Ideone & \href{https://ideone.com/}{https://ideone.com/} \\
        Online Compiler & \href{http://www.onlinecompiler.net/}{http://www.onlinecompiler.net/} \\
        Web Compiler & \href{http://webcompiler.cloudapp.net/}{http://webcompiler.cloudapp.net/} \\
        A+ & \href{https://github.com/Aalto-LeTech/a-plus/}{https://github.com/Aalto-LeTech/a-plus/} \\
        BOSS & \href{https://sourceforge.net/projects/cobalt/}{https://sourceforge.net/projects/cobalt/} \\
        CloudCoder & \href{https://cloudcoder.org/}{https://cloudcoder.org/} \\
        Code Runner for Moodle & \href{https://github.com/trampgeek/CodeRunner}{https://github.com/trampgeek/CodeRunner} \\
        DOMjudge & \href{http://www.domjudge.org}{http://www.domjudge.org} \\
        Mooshak & \href{https://mooshak.dcc.fc.up.pt/}{https://mooshak.dcc.fc.up.pt/} \\
        Online Judge Plugin for Moodle & \href{https://github.com/hit-moodle/moodle-local_onlinejudge}{https://github.com/hit-moodle/moodle-local\_onlinejudge} \\
        SIO2 & \href{https://github.com/sio2project}{https://github.com/sio2project} \\
		TestMyCode & \href{http://testmycode.github.io}{http://testmycode.github.io} \\
        Tsinghua Online Judge & \href{http://dsa.cs.tsinghua.edu.cn/oj/}{http://dsa.cs.tsinghua.edu.cn/oj/} \\
        Virtual programming lab & \href{https://moodle.org/plugins/mod_vpl}{https://moodle.org/plugins/mod\_vpl} \\
        Web-CAT & \href{http://web-cat.org/home}{http://web-cat.org/home} \\
        xLx & \href{http://dbis-group.uni-muenster.de/projects/xlx/}{http://dbis-group.uni-muenster.de/projects/xlx/} \\
        CodeEval & \href{https://www.codeeval.com/}{https://www.codeeval.com/} \\
        Codility & \href{https://codility.com}{https://codility.com} \\
        HackerEarth & \href{https://www.hackerearth.com/}{https://www.hackerearth.com/} \\
        Hackerrank & \href{https://www.hackerrank.com}{https://www.hackerrank.com} \\
        Qualified & \href{https://qualified.io/}{https://qualified.io/} \\       
        \hline
    \end{longtable}
}

\end{document}